\documentclass[letterpaper,journal]{IEEEtran}
\IEEEoverridecommandlockouts
\usepackage{geometry}
\usepackage{url}
\usepackage{amsmath,amsfonts}
\usepackage{algorithm}

\usepackage{algpseudocode}
\usepackage[]{graphicx}
\graphicspath{ {./figs/} }
\usepackage{textcomp}
\usepackage{xcolor}
\usepackage{tikz}
\usetikzlibrary{shapes.geometric,arrows.meta,positioning,fit,backgrounds,calc}

\usepackage{comment}
\usepackage{multirow}
\usepackage{epsfig}
\usepackage{subcaption}

\usepackage{booktabs} 
\usepackage{multirow} 

\usepackage{etoolbox}
\usepackage{array}
\usepackage{tabularx}
\usepackage{hhline}
\usepackage{xspace}
\usepackage{upgreek}
\usepackage{makecell}
\usepackage{etoolbox}
\usepackage{multirow}

\def\BibTeX{{\rm B\kern-.05em{\sc i\kern-.025em b}\kern-.08em
    T\kern-.1667em\lower.7ex\hbox{E}\kern-.125emX}}

\begin{document}


\title{EMSpice 3: Full-chip Temperature-Aware Multiphysics Electromigration and IR-Drop Analysis}

\author{
\indent Haotian Lu~\IEEEmembership{Student Member,~IEEE}, Sheldon X.-D. Tan~\IEEEmembership{Fellow,~IEEE}
~\thanks{
 \indent Haotian Lu and Sheldon X.-D. Tan are with the Department of
 Electrical and Computer Engineering,  University of California,
 Riverside, CA 92521 USA
\newline \indent 
This work is supported in part by NSF grants CCF-2007135 and CCF-2305437.
}
}

\vspace{-0.3in}

\maketitle

\vspace{-0.3in}

\begin{abstract}
   
  This paper presents \textit{EMSpice~3}, a full-chip multiphysics framework for coupled electromigration (EM), thermomigration (TM), and IR-drop analysis of practical power-grid (P/G) networks.
  The framework is, to our knowledge, the first EM-IR analysis flow that jointly incorporates Joule heating and practical spatial thermal profiles for full-chip P/G network designs. 
  It operates on extracted power-grid netlists and combines an immortality check, transient EM/TM stress evolution, void-induced resistance updates, repeated IR-drop recomputation, and optional Monte Carlo lifetime prediction. 
  To make chip-level EM analysis tractable, the framework integrates an extended rational Krylov subspace method into the transient solver, achieving $1.18\times$--$1.50\times$ speedup with 
  sub-0.05\% reported TTF/final-IR metric error relative to the default non-Krylov FDTD analysis across six benchmark designs.
  The numerical results reveal that the specific spatial temperature profile can have a more significant impact on P/G network lifetime than the average temperature itself. 
  In the RISC-V core, a higher-average-temperature profile can avoid the 10\% IR-drop failure threshold when its hotspots are less aligned with critical current paths, while mapped temperature gradients can move the critical void location and change which resistor branches are degraded. 
  Monte Carlo analysis further shows design-specific variation sensitivity: under 20\% variation in 
  EM diffusivity~$\kappa(x)$ and critical stress, the RISC-V core exhibits about 15.8\% TTF coefficient of variation, whereas the ARM Cortex-A logic core exhibits only 0.0058\%.
  These results show that practical thermal profiles, resistance feedback, and stochastic material variation must be considered jointly for predictive full-chip EM-IR analysis.

\end{abstract}


\section{Introduction}
\label{sec:introduction}

Electromigration (EM) remains a major reliability concern in advanced
copper-based VLSI circuits as technology scaling increases current density while
shrinking interconnect dimensions. Traditional single-wire methods such as
Black's equation and Blech's limit are widely used but overly conservative for
complex networks. Physics-based methods that solve Korhonen's equation more
accurately still largely neglect realistic spatial temperature distributions in
both EM degradation and IR-drop analysis.

EM is driven by the momentum transfer from conducting electrons to metal atoms.
As atoms migrate away from the cathode, tensile stress builds there, while
compressive stress accumulates near the anode. This stress gradient opposes the
current-driven atomic flux through stress migration and continues to evolve
until equilibrium is reached. If the tensile stress exceeds a critical
threshold, void nucleation and subsequent growth occur, increasing wire
resistance and eventually causing circuit failure. In addition to
current-induced migration, temperature gradients caused by Joule heating or
active devices can drive atomic transport from hot to cold regions, leading to
thermomigration (TM).

Classical EM models, such as those of Black~\cite{Black:1969fc} and
Blech~\cite{Blech:1976ko}, are therefore limited for modern full-chip
reliability analysis. Korhonen et al.~\cite{Korhonen:1993bb} introduced a
physics-based partial differential equation (PDE) model for stress evolution in
interconnect wires, and a range of analytical and numerical methods have since
been proposed to solve it. For example, Cook et al.~\cite{CookSun:TVSI'18}
proposed \textit{FastEM}, which first discretizes the PDE into linear
time-invariant ordinary differential equations (ODEs) and then applies
Krylov-subspace reduction to accelerate the computation. Chen et
al.~\cite{ChenTan:TCAD'16} developed analytical solutions that provide further
insight into stress evolution in interconnect structures.

Most solutions still overlook spatial temperature gradients. Studies show TM
can be comparable in magnitude to EM~\cite{Abbasinasab:DAC'2018,ChenTan:TCAD'21},
and rising power density and thermal resistance will intensify these
effects~\cite{Todri:TVLSI'13}. Although Joule-heating-aware EM has been
studied~\cite{KavousiChen:ASPDAC'22}, practical chip-level thermal maps are
still rarely incorporated into full-chip EM analysis.

Full-chip reliability analysis therefore requires coupled stress, void, and
IR-drop modeling under realistic thermal fields, not just isolated Korhonen
solves. Prior full-chip EM analysis flows~\cite{SunYu:TDMR'20} moved in this
direction, yet two important gaps remain: (1)~Joule heating and practical
chip-level thermal maps are not fully incorporated together into the EM/IR-drop
loop, even though spatial hotspot alignment strongly affects which trees fail
and when; and (2)~existing approaches offer limited guidance on when
statistical EM analysis is most informative relative to deterministic
prediction.

In this paper, we present a full-chip multiphysics framework for coupled
electromigration (EM), thermomigration (TM), and IR-drop analysis of practical
power-grid interconnects. The framework operates on extracted power-grid
netlists together with material and solver parameters, Joule-heating
information, and practical chip-level thermal maps when available. It combines
an immortality check, transient stress evolution with iterative resistance
feedback, repeated IR-drop recomputation, and optional Monte Carlo analysis for
statistical lifetime prediction. Our numerical results show that realistic
spatial temperature fields can change both the mortal-tree population and the
network time-to-failure even when the average temperature is unchanged. They
also show non-monotonic thermal sensitivity: a higher-average-temperature
profile can be less damaging than a lower-average-temperature profile if its
hotspots do not overlap the critical current paths. Sensitivity to process
variation likewise depends strongly on how close the design is to the
nucleation boundary. These observations motivate a unified analysis flow in
which electrical loading, spatial thermal structure, and stochastic material
variation are treated together. The key contributions of this work are
summarized as follows:

\begin{itemize}

\item We develop a physics-based full-chip framework that couples an
immortality check, transient stress evolution, void-induced resistance updates,
and repeated IR-drop recomputation, enabling network-level EM/TM degradation
analysis rather than isolated-wire lifetime estimation.

\item We incorporate practical spatial temperature maps directly into the coupled
EM/TM/IR-drop analysis. Different thermal maps with the same mean temperature
can produce substantially different mortal-tree counts and can determine
whether the 10\% IR-drop failure threshold is reached within the simulation
horizon. Even peak temperature alone can be misleading when hotspots are
misaligned with the most electrically stressed trees.

\item We observe that the specific spatial temperature profile can have a more
significant impact on network reliability than the average temperature itself:
a higher-average-temperature profile can even yield a better TTF than a lower
average-temperature profile when its hotspots are less aligned with critical
current paths. This non-monotonic behavior is explained by a branch-selection
failure mechanism in the coupled EM/TM and IR-drop analysis: mapped temperature
gradients can move the critical void location within the same tree, changing
which resistor branches are degraded and thereby dominating network-level TTF.
Thus, two profiles with similar average and maximum temperatures can produce
different IR-drop failure outcomes.

\item We integrate an extended rational Krylov acceleration method into the
transient solver with three implementation optimizations, achieving
$1.18\times$--$1.50\times$ speedup across six industrial-scale designs with
0.0\% reported TTF/final-IR metric error relative to the default non-Krylov
FDTD analysis, thereby making the coupled transient analysis substantially more
tractable on large trees.

\item We show that process-variation sensitivity is strongly regime-dependent.
The RISC-V core (18 mortal trees) exhibits $\approx$15.8\% TTF coefficient of
variation (CoV) under 20\% variation in
EM diffusivity~$\kappa(x)$ and critical
stress, while the ARM Cortex-A logic core (206 mortal trees under the same
353\,K Joule-heating thermal condition) shows only
$\approx$0.0058\%, clarifying when statistical EM analysis materially changes the
predicted lifetime distribution.

\item We validate the framework on six industrial-scale designs, demonstrating
that the immortality check avoids unnecessary transient simulation while
the combined modeling and acceleration strategy remains tractable at full-chip
scale.
    

\end{itemize}

The remainder is organized as follows. Section~\ref{sec:related} reviews
related work; Section~\ref{sec:review} covers the multiphysics background;
Section~\ref{sec:emspice3_framework} describes the proposed framework and
acceleration strategy; Section~\ref{sec:observations} presents numerical
results; and Section~\ref{sec:conclusion} concludes.

\section{Review of related works}
\label{sec:related}

Early EM work relied on empirical models such as those of
Black~\cite{Black:1969fc} and Blech~\cite{Blech:1976ko}, which are still widely
used but overly conservative for network-level analysis and do not account for
realistic thermal effects.

Korhonen et al.~\cite{Korhonen:1993bb} introduced a physics-based PDE model for
stress evolution in interconnect wires, enabling more accurate EM analysis at
the segment level. As VLSI technology has advanced, however, it has become
increasingly important to incorporate both electromigration and
thermomigration, because power density and thermal resistance continue to rise
in advanced ICs.

The role of temperature in EM has historically been underestimated, and many
models still fail to capture dynamic thermal conditions across an operating
chip. Studies by Chen and Tan~\cite{ChenTan:TCAD'21} and Abbasinasab et
al.~\cite{Abbasinasab:DAC'2018} highlight the importance of accurate
temperature modeling for wire mortality and metal migration. More recent EM/TM work, including
that of Kavousi and Chen~\cite{KavousiChen:ASPDAC'22} and
Lamichhane et al.~\cite{Lamichhane:ISVLSI'24}, moves toward temperature-aware EM
analysis, but practical chip-level thermal maps remain only partially
integrated into the reliability loop.

More recently, the open-source EM analysis tool PROTON was proposed
in~\cite{Olympia:SMACD'23}. Like \textit{EMSpice}, PROTON uses finite differences for
stress analysis, but it relies on an external commercial simulator for IR-drop
analysis and therefore does not capture the same integrated EM/IR-drop
interaction. In addition, PROTON does not include an immortality screen to
filter out EM-immune wires before transient simulation. Overall, while these
tools offer more realistic EM analysis than classical empirical models, they
still do not fully account for Joule heating and the resulting thermomigration
effects in both screening and transient stress analysis.

\section{Multiphysics modeling for EM and IR-drop analysis}
\label{sec:review}
\subsection{Temperature modeling}
The importance of temperature in electromigration (EM) analysis has
historically been underestimated. As very-large-scale integration (VLSI)
technology continues to scale, however, temperature has become a critical
factor in metal migration~\cite{ChenTan:TCAD'20,Abbasinasab:DAC'2018}.
Accurate temperature modeling is therefore essential for realistic prediction
of wire mortality and stress evolution.
Starting from the first law of thermodynamics, the time-dependent 
thermal behavior in solids is described by the following equation~\cite{carslaw_conduction_1959}:
\begin{equation} 
    \small
    \begin{aligned} 
        \rho_m c_p \frac{\partial T}{\partial t} + \nabla q = Q 
    \end{aligned} 
    \label{eq:thermal_equation} 
\end{equation} 
Here, $c_p$ represents the specific heat capacity, $\rho_m$ denotes 
the mass density, $q=-k\nabla T$ is the heat flux defined by Fourier's 
law, and $Q = Q_{jh} + Q_{conn}$ encompasses the components of Joule 
heating and heat convection, where $Q_{jh}=j^2\rho$ and 
$Q_{conn}=k(T-T_0)/\Gamma^2$. The thermal characteristic length, 
$\Gamma$, depends on the geometry and topology of the environment, and 
it is calculated as~\cite{Abbasinasab:DAC'2018}:
\begin{equation} 
    \small
    \begin{aligned} 
        \Gamma^2 = t_{cu}t_{ILD}\frac{k_{cu}}{k_{ILD}} 
    \end{aligned} 
\end{equation} 
In this context, $t_{cu}$ and $k_{cu}$ refer to the thickness and 
thermal conductivity of the wire, respectively, and $t_{ILD}$ 
and $k_{ILD}$ refer to the thickness and conductivity of the dielectric.

Under stationary conditions, Eq.~\eqref{eq:thermal_equation} simplifies to: 
\begin{equation} 
    \small
    \begin{aligned} 
        k\nabla^2T - (j^2\rho+k(T_0-T)/\Gamma^2) = 0 
    \end{aligned} 
    \label{eq:stationary_thermal_equation} 
\end{equation} 
Eq.~\eqref{eq:stationary_thermal_equation} can be solved numerically by
finite differences (FDM) or finite elements (FEM) to obtain the spatial
temperature distribution.
The temperature profile along a single wire, subject to boundary 
conditions $T(-L/2)=T_1$ and $T(+L/2)=T_2$, is given 
by~\cite{Abbasinasab:DAC'2018,KavousiTan:ICCAD'20}: 
\begin{equation} 
    \small
    \begin{aligned} T(x) = T_0 + T_m \left[1-\frac{\cos\left(\frac{x}{\Gamma}\right)}{\cos\left(\frac{L}{2\Gamma}\right)}\right] + T_n \left[1-\frac{\sin\left(\frac{x}{\Gamma}\right)}{\sin\left(\frac{L}{2\Gamma}\right)}\right] 
    \end{aligned} 
    \label{eq:temp_solution} 
\end{equation}
Here, $T_m = j^2\rho\Gamma^2/k_{cu}$ is the symmetric (Joule-heating)
temperature rise and $T_n = (T_1 - T_2)/2$ is the antisymmetric
(boundary-asymmetry) component.
Prior approaches often assume static boundary temperatures $T_1$ and $T_2$
for the whole chip~\cite{KavousiChen:ASPDAC'22, Lamichhane:ISVLSI'24},
which fails to capture the spatially varying thermal landscape of an
operating die. \textit{EMSpice~3} retains the model of
Eq.~\eqref{eq:temp_solution} but computes boundary conditions per tree.
When an external thermal map is provided, via-node temperatures are read
from the map and interpolated to each tree's endpoints; unconstrained
endpoint nodes are solved in tridiagonal thermal blocks. Without an
external map, the tool falls back to Joule-heating-only mode using the
ambient reference $T_0$. After per-tree endpoints are resolved,
Eq.~\eqref{eq:temp_solution} is evaluated along every branch under the
local current density, yielding a full per-tree temperature profile for
downstream TM-aware screening and transient simulation.

\subsection{Steady-state-stress-based screening}
In \textit{EMSpice~3}, the primary immortality screen is a one-shot
steady-state stress analysis on the original netlist. This screening step is
optional and is performed before the transient EM/TM simulation begins. Its
purpose is not to declare failure directly, but to identify the wires and
trees that require transient follow-up.

As shown later in Section~\ref{sec:observations}, this screening approach does
not account for the full interaction among trees over the entire EM aging
process. As a result, it can identify only a subset of the truly immortal
wires. Any wire that is not screened out still requires transient follow-up to
determine its actual mortality. Even so, the screening remains useful in design
exploration because it removes clearly immortal trees before the expensive
transient stage.

For the non-TM-aware case, the screening flow first extracts the trees from
the original power-grid netlist and runs the IR-drop solve to obtain the
branch current densities. The original tree geometry is then combined
with these current densities to build the steady-state tree set. For
each tree, \textit{EMSpice~3} solves the steady-state form of the
stress equation and samples the resulting hydrostatic stress on the same
branch-local stress grid used by the transient solver. A branch is
classified as requiring transient follow-up when any finite sampled node
stress on that branch exceeds the critical stress
$\sigma_{\mathrm{crit}}$. Otherwise, the branch is classified as
immortal. A tree requires transient follow-up when at least one of its
branches does; otherwise, the entire tree is screened out as immortal.

For the TM-enabled case, the flow first performs the one-time thermal
analysis described above. The resulting
temperature field is then used in the same steady-state stress solve,
now with thermomigration-aware matrix stamping. The classification rule
remains unchanged: branches and trees are screened according to whether
their maximum finite steady-state stress exceeds
$\sigma_{\mathrm{crit}}$. In this sense, the TM flow differs only in
the thermal field that enters the steady-state stress solve; the
screening output is still reported as immortal wires/trees and
wires/trees that require transient analysis.

This steady-state criterion is consistent with the implementation in
\textit{EMSpice~3} because it directly measures the stress state that
drives void nucleation, while avoiding the need to infer mortality from
an intermediate voltage criterion. The screen therefore acts as a
physics-based preclassification step that reduces the transient workload
without labeling all non-immortal branches as already failed.

\subsection{IR-drop in VLSI interconnects}
IR-drop is increasingly important in modern designs because rising wire
resistance degrades on-chip power delivery. As fabrication technologies shrink
and current density increases, the dimensions of the power-carrying wires also
decrease. In the context of EM analysis, which evolves over days or months, the
electrical potential distribution can be treated as
quasi-static~\cite{SunYu:TDMR'20}. It is therefore governed by the strong form
of the Laplace equation:
\begin{equation}
    \small
    \nabla \cdot \left(\frac{1}{\rho(\phi)} \nabla u \right) = 0 \quad \text{in } \Omega_L
    \label{eq:ir_drop_laplace}
\end{equation}
\begin{equation}
    \small
    u = g_u \quad \text{on } \partial \Omega_L \cap \Gamma_u
\end{equation}
\begin{equation}
    \small
    \vec{n} \cdot \frac{1}{\rho(\phi)} \nabla u = g_j \quad \text{on } \partial \Omega_L \cap \Gamma_j
\end{equation}

where $\rho(\phi)$ represents the copper resistivity influenced by the phase
field. $\Gamma_u$ denotes boundaries with Dirichlet voltage conditions and
$g_u$ is the imposed voltage source. $\Gamma_j$ denotes boundaries with
Neumann current-flux conditions and $g_j$ is the corresponding current-density
source. For full-chip EM analysis, directly solving
Eq.~\eqref{eq:ir_drop_laplace} is computationally expensive. Instead, we assume
uniform current density along each wire segment, extract a resistance network
from the power-grid layout, and apply modified nodal analysis (MNA), which
gives
\begin{equation}
    \small
    M(t) u(t) = P I(t) 
    \label{eq:ir_drop_mna}
\end{equation}
Here, $M(t)$ denotes the power-grid admittance matrix, which changes over time
as wire resistances degrade due to EM. The matrix $P \in \mathbb{R}^{b\times p}$
maps the $p$ current sources $I(t)$ to the nodal-voltage vector $u(t)$. The
resulting nodal voltages and branch currents are then used in both the
temperature-aware immortality screen and the transient EM/TM stress solve.

\subsection{EM stress evolution mechanism}
In copper dual damascene interconnects, the electron wind force drives
atoms from the cathode toward the anode. Because the wire is encased in
a rigid dielectric, this redistribution generates tensile hydrostatic
stress at the cathode and compressive stress at the anode. Once the
tensile stress exceeds a critical threshold, a void nucleates.
Physics-based models~\cite{TanAmrouch:2017int,TanTahoori:Book'19}
characterize the subsequent EM failure process through three sequential
phases: nucleation ($t_{\mathrm{nuc}}$), incubation ($t_{\mathrm{inc}}$), and growth
($t_{\mathrm{growth}}$).
During nucleation, stress accumulates, potentially leading to void 
formation at critical stress levels at cathode nodes. In the incubation 
phase, void growth occurs without resistance changes. The growth phase 
begins when the void reduces the cross-sectional area of the via, 
causing current to pass through a higher resistive barrier, changing 
the wire resistance. This phase may result in early failure 
(open circuit) or progressive resistance increase up to a predefined 
threshold (e.g., 10\% change). The total time to failure (TTF) is the
sum of these three phases:
\begin{equation}
    \small
    TTF = t_{\mathrm{nuc}} + t_{\mathrm{inc}} + t_{\mathrm{growth}}
    \label{eq:ttf_calculation}
\end{equation}

\subsubsection{Void nucleation phase}
The nucleation phase of the electromigration-thermomigration 
(EM/TM) process marks the initial period up to the point where 
a void forms within the interconnect structure being analyzed. 
This phase begins with an inherent residual stress within the
system.

The evolution of EM stress, 
$\sigma$, driven by the combined effects of electromigration 
(EM), stress migration (SM), and thermomigration (TM), is 
effectively modeled through the application of Korhonen's 
equation~\cite{DEORIO:Micro'2010}:
\begin{equation} 
    \small
    \begin{aligned} 
        \frac{\partial\sigma}{\partial t} = \frac{\partial}{\partial x}\left[\kappa(x)\left(\frac{\partial\sigma}{\partial x}-\frac{eZ\rho j}{\Omega}-\frac{Q}{\Omega T}\frac{\partial T}{\partial x}\right)\right]
    \end{aligned} 
\end{equation} 
Here, $\kappa(x) = D_a(T(x))B\Omega / (k_B T(x))$ represents diffusivity,
where $D_a = D_0 \exp(-E_a/k_BT)$ is the atomic diffusion coefficient,
$D_0$ a constant, and $E_a$ the activation energy for EM.
The transient hydrostatic stress evolution, $\sigma(x,t)$, during the
nucleation phase of a general interconnect structure is governed by the
Korhonen equation:
\begin{equation} 
    \small
    \begin{aligned} \text{PDE: } \frac{\partial\sigma}{\partial t} & = \frac{\partial}{\partial x}\left[\kappa(x)\left(\frac{\partial\sigma}{\partial x}-S-M\right)\right], \quad t > 0 
    \end{aligned} 
    \label{eq:tm_aware_korhonen_pde} 
\end{equation} 
\begin{equation} 
    \small
    \begin{aligned} \text{BC: } \kappa(x_b)\left(\left.\frac{\partial\sigma}{\partial x}\right|_{x=x_b} - S-M\right) & = 0, \quad 0 < t < t_{\text{nuc}} 
    \end{aligned} 
    \label{eq:tm_aware_korhonen_bc} 
\end{equation} 
\begin{equation} 
    \small
    \begin{aligned} \text{IC: } \sigma(x,0) = \sigma_T 
    \end{aligned} \label{eq:tm_aware_korhonen_ic} 
\end{equation} 
In these equations, $S = \frac{eZ\rho j}{\Omega}$ represents the EM 
flux and $M = \frac{Q}{\Omega T}\frac{\partial T}{\partial x}$ the TM 
flux. $\kappa(x) = D_a(T(x))B\Omega/(k_B T(x))$ indicates 
position-dependent diffusivity due to non-uniform temperature, 
with $D_a = D_0 \exp(-E_a/(k_B T(x)))$ as the atomic diffusion rate, 
$\sigma_T$ the initial thermally induced residual stress, and $x_b$ 
marking the block terminals.

\subsubsection{Void incubation phase}
\begin{figure}[!htb]
    \centering
    \includegraphics[width=0.58\columnwidth]{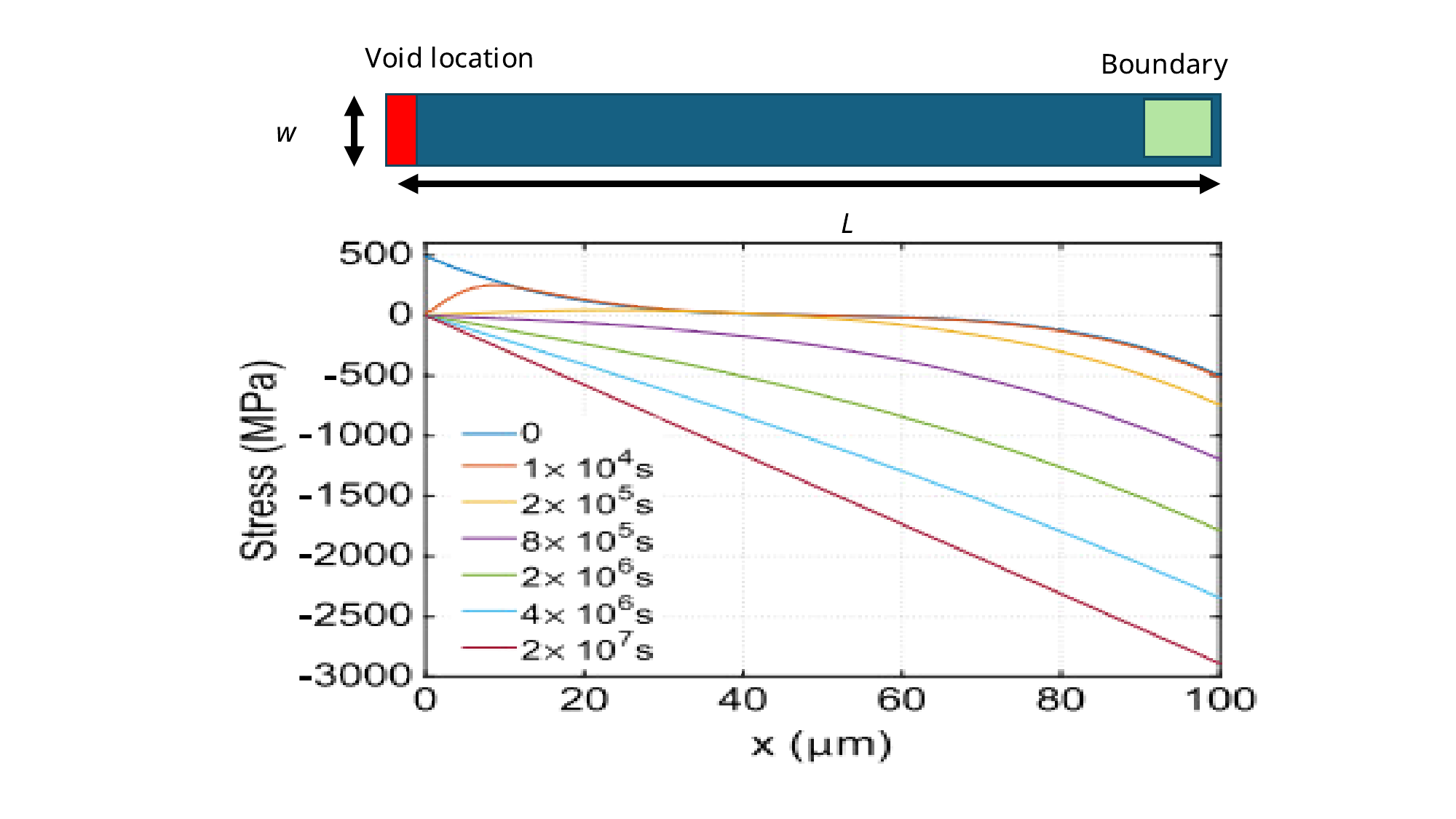}
    \caption{\small Stress distribution with void}
    \label{fig:void_stress_dist}
\end{figure}

When the stress reaches a critical threshold at $t_{\mathrm{nuc}}$, a void
forms. Nevertheless, the interconnect resistance remains
largely unchanged because the void has not yet consumed the
critical volume of the via cross-section. Typically, the void 
nucleates at or near a terminal node when stress reaches this 
critical point. At this juncture, the normal component of stress 
at the void boundary is generally zero as shown in 
Fig.~\ref{fig:void_stress_dist}. 
To model this, we consider the 
effective thickness of the copper-void boundary, $\delta$, as 
infinitesimally small. We then establish the stress gradient 
between the zero-stress void surface and the adjacent metal as 
indicated in~\cite{Korhonen:1993bb, sukharev2016postvoiding}:
\begin{equation}
    \small
    \frac{\partial \sigma}{\partial x} \Bigg|_{x=x_{\text{nuc}}} = \frac{\sigma(x_{\text{nuc}}, t)} {\delta} , \quad t_{\text{nuc}} < t < \infty
\label{eq:incubation_bc}
\end{equation}
In the above expression, $x_{nuc}$ represents the location where 
the void nucleates along the boundary, and $\delta$ denotes the 
thickness of the void interface. The volume of the void, 
$V_v(t)$, in a multi-segment wire is calculated based on the 
stress distribution throughout the remaining wire sections. 
This volume satisfies the atom conservation equation, which is 
expressed as follows:
\begin{equation}
    \small
    V_v(t) = -\int_{\Omega_L} \frac{\sigma(t)}{\mathcal{B}} \, dV
    \label{eq:void_volume_equation}
\end{equation}
where $\mathcal{B}$ is the effective bulk elastic modulus of the wire.
This relationship is derived from the principles of atom
conservation across different segments of the wire as outlined in~\cite{Korhonen:1993bb, sukharev2016postvoiding}.
Here $\Omega_L$ represents the volume of the remaining interconnect 
wire, and $V$ denotes the total volume of the wire, defined by
$V = W \times h \times l$, where $W$ is the width
of the wire, $h$ is its thickness, and $l$ is its length. It 
is important to note that Eq.~\eqref{eq:void_volume_equation}
applies equally during the void growth phase.

The period of incubation is defined by the interval
$t_{\mathrm{nuc}} < t < t_{\mathrm{inc}}$,
where the void volume $V_v(t)$ remains below the critical volume 
$V_{\text{crit}}$:
\begin{equation}
    \small
    V_v(t) < V_{\text{crit}} \quad
    \label{eq:incubation_phase_volumes}
\end{equation}
In this context, $V_{\text{crit}}$ is dependent primarily on the 
cross-sectional area of the wire. In a one-dimensional scenario, 
the width of the wire $W$ is used as the critical dimension
rather than the volume, marking the end of the incubation
phase as resistance begins to change. This usage arises 
because $W \geq h$, and the void will consume the entire 
cross-section of the wire once it expands to the width $W$. 
Fig.~\ref{fig:void_stress_dist} illustrates the stress dynamics in 
a wire during the incubation phase. The stress at the cathode 
rapidly drops to zero, while the anode stress gradually declines 
until it stabilizes. This decline fosters void growth, driven by
the accumulation of negative stress as described by
Eq.~\eqref{eq:incubation_bc}. The void volume reaches saturation 
when the stress stabilizes. 

During the incubation phase, the void volume remains insufficient 
to obstruct the via's cross section and stays below the critical 
volume. At the termination of the incubation phase, a void will 
encompass the entire cross section of the via, potentially 
leading to either an early or a late failure of the 
wire~\cite{Alam:2007}. Early failures are common in structures where 
vias are placed above lines. Here, the void obstructs the via, 
severing the connection to the upper layer, which consists of 
non-conductive materials like $\mathrm{Si_3N_4}$. This results in a 
failure at the close of the incubation phase. Conversely, 
late failures occur in structures with vias below the lines. 
In such cases, even when a void reaches a critical size within a 
via-below line, current may still traverse the barrier layer, 
leading to an increase in resistance during the growth phase
~\cite{SunYu:TDMR'20}.

\subsubsection{Void growth phase}
During the void growth phase, the void continues to expand and wire
resistance begins to change measurably, distinguishing this phase from
the incubation phase. This resistance increase arises because the
electrical current is forced to traverse a highly resistive barrier
layer, often composed of Ta/TaN, whose resistivity is significantly
greater than that of Cu. The change in wire resistance during this
phase can be approximately expressed as~\cite{SunYu:TDMR'20}:
\begin{equation}
    \small
    \begin{aligned}
        \Delta R(t) = \frac{V_{\nu}-V_{crit}}{WH}\left[\frac{\rho_{T_a}}{h_{T_a}(2H+W)}-\frac{\rho_{C_u}}{HW}\right]
    \end{aligned}
\end{equation}
Here, $\rho_{T_a}$  and $\rho_{C_u}$ represent the resistivities of the 
barrier material (Ta/TaN) and copper, respectively. 
$W$ denotes the width of the wire, $H$ refers to the thickness 
of the copper, and $h_{Ta}$  is the thickness of the barrier layer.

\subsubsection{Finite-difference transient formulation}
Korhonen's equation, including the EM, SM, and TM terms in
Eq.~\eqref{eq:tm_aware_korhonen_pde}, is solved by a finite-difference
transient formulation on each multi-segment interconnect tree, similar
to~\cite{SunYu:TDMR'20}. After finite-difference discretization, the PDE in
Eq.~\eqref{eq:tm_aware_korhonen_pde}, together with its boundary conditions
(BCs) and initial condition (IC), is transformed into the linear descriptor
system
\begin{equation}
    \small
    \begin{aligned}
        C\dot{\boldsymbol{\sigma}}(t) & = A\boldsymbol{\sigma}(t) + B\boldsymbol{j}(t) - D \\
        \boldsymbol{\sigma}(0) & = [\sigma_1(0), \sigma_2(0), \ldots, \sigma_n(0)]^{\top}
    \end{aligned}
\label{eq:em_tm_ode}
\end{equation}
In this formulation, $\boldsymbol{\sigma}(t)$ denotes the stress vector; $C$ and $A$ are
$n \times n$ matrices; $D$ is an $n \times 1$ vector associated with the TM
term; and $B$ is an $n \times p$ input matrix, where $p$ is the number of
current-density inputs. The initial condition $\boldsymbol{\sigma}(0)$ includes both the
residual stress and, at later outer steps, the stress carried over from the
preceding simulation state. Combining Eq.~\eqref{eq:ir_drop_mna},
Eq.~\eqref{eq:void_volume_equation}, and Eq.~\eqref{eq:em_tm_ode}, the coupled
EM/TM/IR-drop system addressed in this work can be summarized as
\begin{equation}
    \small
    \begin{aligned}
        &C\dot{\boldsymbol{\sigma}}(t)  = A\boldsymbol{\sigma}(t) + B\boldsymbol{j}(t) - D \\
        &V_{v}(t) = -\int_{\Omega_L}^{}\frac{\sigma(t)}{\mathcal{B}} dV \\
        &M(t)u(t) = P I(t) \\
        &\boldsymbol{\sigma}(0)  = [\sigma_1(0), \sigma_2(0), \ldots, \sigma_n(0)]^{\top}
    \end{aligned}
\label{eq:em_tm_ir_ode}
\end{equation}
Eq.~\eqref{eq:em_tm_ir_ode} summarizes the multiphysics system used to
analyze coupled IR-drop, void volume, and stress evolution over time. The
implementation details are presented in Section~\ref{sec:emspice3_framework}.

\section{Proposed EM/TM-aware IR-drop Analysis Framework with Accelerated Strategy}
\label{sec:emspice3_framework}

\subsection{Overall analysis flow}
\begin{figure}[!htb]
  \centering
  \includegraphics[width=0.92\linewidth]{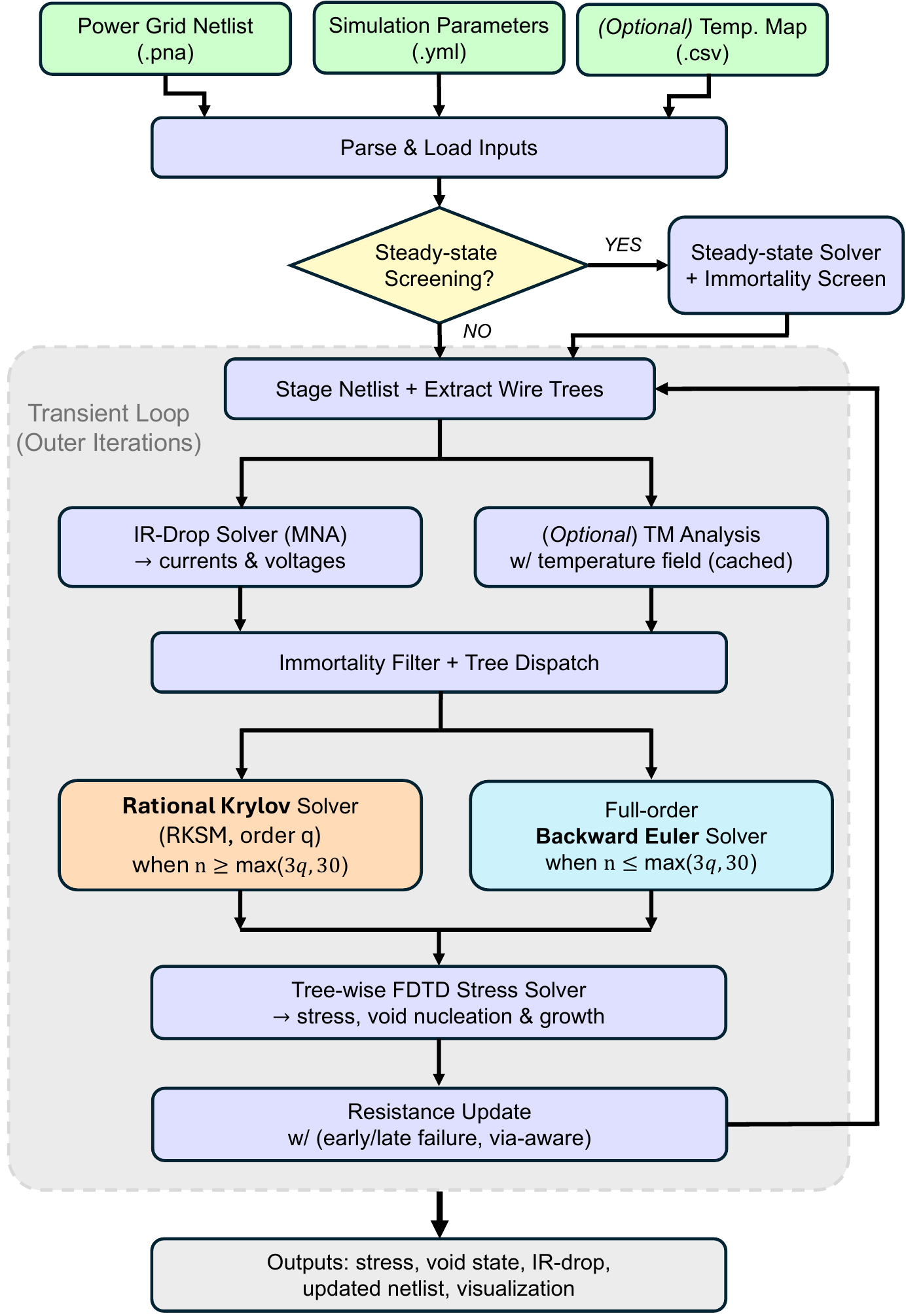}
  \caption{\small Overall algorithmic flow of the proposed \textit{EMSpice~3} framework}
  \label{fig:emspice3_framework}
\end{figure}

Fig.~\ref{fig:emspice3_framework} summarizes the overall analysis flow as
implemented in \textit{EMSpice~3}. The analysis starts from three inputs: an
extracted power-grid netlist, a YAML parameter file containing the material and
numerical settings, and an optional thermal map for TM-aware analysis. After
parsing these files, the framework can perform an optional steady-state
screening pass to prune trees that are already classified as immortal.

At each aging step the degraded netlist is partitioned into wire trees; MNA
solves the IR-drop problem for nodal voltages and branch current densities,
while the thermal field (when enabled) is loaded once and cached. Larger trees
are accelerated by the rational Krylov path; smaller trees use backward Euler (BE)
directly. Updated stress and void states are converted to resistance changes
and written back to the netlist for the next outer step, closing the coupled
electrical-thermal-reliability loop.

\subsection{Finite-difference transient stress solver}
The finite-difference transient stress solver combines the current-density data
from the IR-drop solution with the temperature distribution to solve the
discretized Korhonen equation over time. For each tree, the solver updates the
hydrostatic stress state, tracks void nucleation and growth, and computes the
resulting change in resistance. These resistance updates are then written back
to the staged netlist so that the next IR-drop solve reflects the current aging
state of the power grid.

\subsection{Early failure detection with via-aware resistance updates}
Once the FDTD solver has advanced the stress state and updated void positions
and lengths, the resulting resistance changes must be written back into the
staged netlist before the next IR-drop solve. \textit{EMSpice~3} distinguishes
two physically distinct failure modes depending on whether the cathode node of
the affected tree carries an explicit via connection to a higher metal layer.

In the common case, \emph{late failure}, the void grows within a wire
segment but does not fully block an inter-layer current path.  The resistance
increase is computed incrementally from the fraction of the wire cross-section
consumed by the void, and the branch resistance is updated accordingly.  When
the cathode node of the tree is connected upward through an explicit via
resistor to a strictly higher metal layer, however, the growing void can sever
the only current path through that via, causing an abrupt open-circuit rather
than a gradual degradation.  This condition is termed \emph{early failure},
and the affected branch resistance is set to an open-circuit sentinel value.

Correctly identifying early failure requires explicit via information.
\textit{EMSpice~3} retains all via resistances during PNA-to-SPICE conversion,
then parses the netlist for two-terminal resistors connecting nodes on
different metal layers. For each tree, the cathode node is resolved from the
topology and its upward-layer connectivity is looked up in the resulting via
map.
In contrast, earlier \textit{EMSpice~1/2}~\cite{SunYu:TDMR'20, Lamichhane:ISVLSI'24} dropped vias
during conversion and therefore could not distinguish the two failure modes.
\textit{EMSpice~3} falls back to late-failure treatment when via resistors are
absent, preserving backward compatibility.


\subsection{Treatment of the void nucleation on mid-tree junction nodes}

\begin{figure}[!htb]
  \centering
  \begin{subfigure}[b]{0.48\columnwidth}
    \centering
    \includegraphics[width=\linewidth]{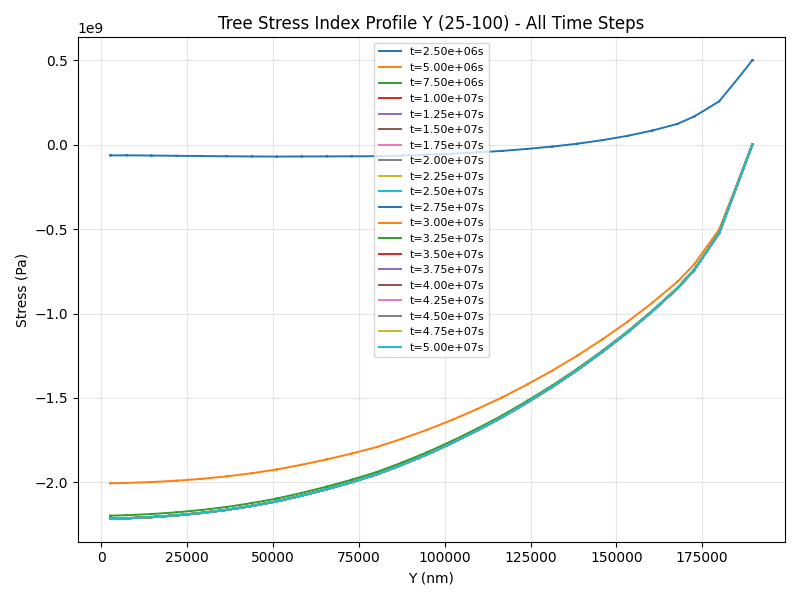}
    \caption{\small Spatial stress profile}
    \label{fig:mid_tree_profile}
  \end{subfigure}
  \hfill
  \begin{subfigure}[b]{0.48\columnwidth}
    \centering
    \includegraphics[width=\linewidth]{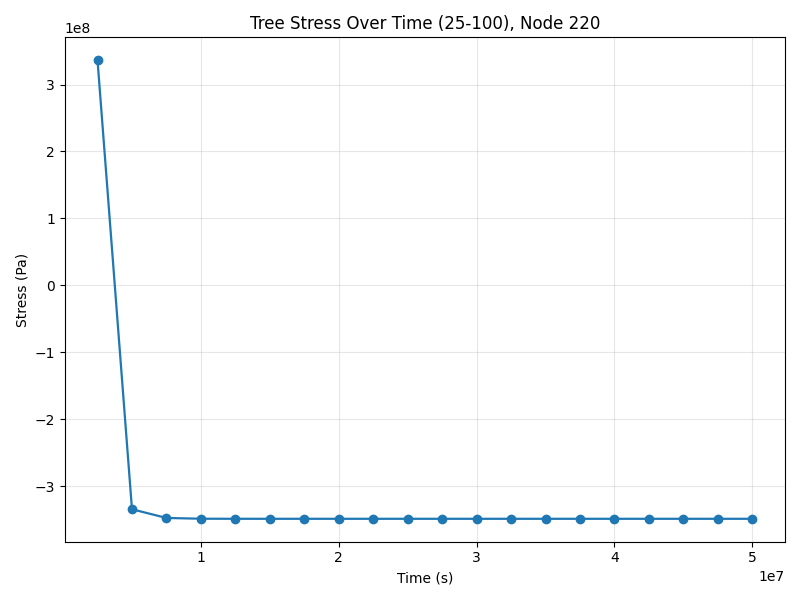}
    \caption{\small Junction-node stress history}
    \label{fig:mid_tree_node}
  \end{subfigure}

  \caption{\small Mid-tree void nucleation in the Dual-Port RAM design.
  Left: stress profile along tree (25-100) over outer timesteps from
  $2.5\times10^6$\,s to $5.0\times10^7$\,s, showing the evolution from a
  near-zero state to a strong compressive gradient as degradation accumulates.
  Right: stress history at junction node 220, where the abrupt transition from
  about $3.3\times10^8$\,Pa tensile stress to about $-3.4\times10^8$\,Pa
  compressive stress marks void nucleation; the subsequent saturation confirms
  that the Robin boundary condition enforces flux decoupling at the interior
  node without dynamic tree splitting.}
  \label{fig:mid_tree_nucleation}
\end{figure}


In existing EM simulation frameworks based on the extended Korhonen
formulation, void nucleation at an interior node of a wire tree is
handled by explicitly splitting the tree into two separate sub-trees at
the nucleation point~\cite{Torosyan:JVST2020}.  A
zero-flux (homogeneous Neumann) boundary condition is then imposed on
each sub-tree at the split boundary, reflecting the physical fact that
the void surface carries no net atomic flux.  While this approach is
physically correct, it requires dynamic bookkeeping: the tree topology
must be restructured at every nucleation event, and the system matrices
must be reassembled for the resulting sub-trees at each subsequent time
step.

\textit{EMSpice~3} avoids tree splitting entirely through a direct
application of the Robin boundary condition at the void node.  Once a
void nucleates at stress-grid node $v_i$, the corresponding row of the
assembled stiffness matrix is cleared and replaced with a Robin stencil
stamped over every segment incident to $v_i$:
\begin{equation}
  \small
    \frac{\partial\sigma}{\partial x}\bigg|_{v_i}
    = \frac{\sigma(v_i,\,t)}{\delta}, \quad t > t_{\mathrm{nuc}},
    \label{eq:robin_mid_tree}
\end{equation}
where $\delta$ is the infinitesimally thin void-surface thickness
already used in the incubation-phase boundary condition
(Eq.~\eqref{eq:incubation_bc}).  Because the stamping loop iterates over
all incident segments regardless of their count, the same code path
handles terminal nodes, junction nodes, and arbitrary mid-tree interior
nodes without any special casing.  As $\delta \to 0$ the large diagonal
coefficient $\kappa/(\Delta x \cdot \delta)$ forces the atomic flux to
zero across all branches meeting at $v_i$, effectively decoupling the
two sub-trees within a very short simulation time.  The decoupling is
therefore not instantaneous but is completed in a time period
proportional to $\delta$, which is kept negligibly small relative to
the simulation time step.

This treatment significantly simplifies the implementation: no tree
topology changes are performed at runtime, and the matrix assembly
routine requires only a localized row replacement rather than a global
remeshing.  As illustrated in Fig.~\ref{fig:mid_tree_nucleation}, the
Robin-based approach reproduces the same physical boundary condition as
the split-tree method while preserving the correctness of the EM stress
dynamics.

\subsection{Rational Krylov subspace acceleration}
\label{subsec:rakrylov}

For power-grid trees with a large number of finite-difference stress nodes, the
backward Euler solve dominates the per-timestep cost of EM/TM analysis.
\textit{EMSpice~3} therefore adopts the recently proposed extended rational
Krylov subspace method ({\it ExtRaKrylovEM})~\cite{LuTan2026RationalKrylovEM},
which projects the $n$-dimensional stress system onto a much smaller subspace
of order $q$ ($q \ll n$). This reduces the dominant linear-algebra cost on
large trees while preserving high solution accuracy. Compared with the standard
extended Krylov method, {\it ExtRaKrylovEM} uses a shifted resolvent operator
to construct the projection basis, which better captures the transient dynamics
of interest and avoids numerical issues associated with near-singular system
matrices under Neumann-type boundary conditions.

\subsubsection{System formulation and extended rational Krylov subspace}

After finite-difference spatial discretization of the Korhonen equation, the
per-tree stress evolution is described by the linear time-invariant system
\begin{equation}
  \small
    C\dot{\boldsymbol{\sigma}}(t)
    = A\,\boldsymbol{\sigma}(t) + B\,\boldsymbol{j} - D,
    \qquad \boldsymbol{\sigma}(0) = \boldsymbol{\sigma}_0,
    \label{eq:rak_lti}
\end{equation}
which is the same descriptor form as Eq.~\eqref{eq:em_tm_ode}. In this paper,
$\boldsymbol{j}$ is assumed constant over one outer \textit{EMSpice~3} update,
but the thermomigration term satisfies $D \neq 0$ in general. Define the
constant forcing vector
\[
    \boldsymbol{f} := B\,\boldsymbol{j} - D.
\]
Applying the Laplace transform gives
\[
    (sC-A)\,\boldsymbol{\Sigma}(s)
    = C\boldsymbol{\sigma}_0 + \frac{\boldsymbol{f}}{s}.
\]
Because the input is constant, define
$\widetilde{\boldsymbol{\Sigma}}(s) := s\,\boldsymbol{\Sigma}(s)$, so that
\[
    (sC-A)\,\widetilde{\boldsymbol{\Sigma}}(s)
    = sC\boldsymbol{\sigma}_0 + \boldsymbol{f}.
\]
We expand the solution around a shifted frequency $s_0 = 1/t_s$, where $t_s$
is a user-supplied shift time (time constant of interest). Define
\[
    K_{s_0} := s_0 C - A,\qquad R_{s_0} := K_{s_0}^{-1},
\]
and let $s=s_0+\Delta$. Expanding
\[
    \widetilde{\boldsymbol{\Sigma}}(s_0+\Delta)
    = \sum_{k=0}^{\infty}\widetilde{\boldsymbol{m}}_k \Delta^k
\]
and matching powers of $\Delta$ yields the compact recurrence
\begin{equation}
  \small
    \widetilde{\boldsymbol{m}}_k =
    \begin{cases}
        R_{s_0}\!\left(s_0 C\boldsymbol{\sigma}_0 + \boldsymbol{f}\right),
        & k = 0,\\[2pt]
        R_{s_0}\!\left(C\boldsymbol{\sigma}_0 - C\widetilde{\boldsymbol{m}}_0\right),
        & k = 1,\\[2pt]
        -R_{s_0}C\,\widetilde{\boldsymbol{m}}_{k-1},
        & k \ge 2.
    \end{cases}
    \label{eq:rak_moment_recurrence}
\end{equation}
Since $\boldsymbol{j}$ is constant here, no source-moment matching is needed;
all higher-order shifted moments propagate only through the operator
$R_{s_0}C$. This leads to the starting block
\begin{equation}
  \small
    \mathcal{B}_{s_0}
    := \bigl[\,R_{s_0}C\boldsymbol{\sigma}_0,\; R_{s_0}\boldsymbol{f}\,\bigr].
    \label{eq:rak_start_block}
\end{equation}
All shifted response moments therefore lie in the \emph{block rational Krylov
subspace}
\begin{equation}
  \small
    \begin{aligned}
    \mathcal{K}_q\!\left(R_{s_0}C,\,\mathcal{B}_{s_0}\right)
    &= \mathrm{range}\!\left(\bigl[\,
      \mathcal{B}_{s_0},\;
      (R_{s_0}C)\mathcal{B}_{s_0},\;
      \ldots,\right.\\
    &\qquad\left.(R_{s_0}C)^{q-1}\mathcal{B}_{s_0}
    \,\bigr]\right).
    \end{aligned}
    \label{eq:rak_krylov_subspace}
\end{equation}
An orthonormal projection basis $V_r \in \mathbb{R}^{n \times r}$ is
constructed from this subspace via a block Arnoldi process up to order $q$, as summarized
in Algorithm~\ref{alg:emspice_rakrylov}.  Because $s_0$ is chosen to
match the dominant time scale of the simulation, the resulting basis
captures the transient dynamics more accurately than a standard (unshifted)
Krylov expansion for the same order $q$.  Furthermore, computing the
resolvent $(s_0 C - A)$ rather than $A^{-1}$ directly avoids the
near-singularity issue of $A$ under Neumann-type boundary conditions that
has been reported for earlier Krylov EM solvers~\cite{Chatterjee:2016ICCAD}.

\begin{algorithm}[t]
\caption{Constant-Input Block Rational Krylov Reduction}
\label{alg:emspice_rakrylov}
\small
\begin{algorithmic}[1]
\Require System matrices $A$, $C$, forcing vector $\boldsymbol{f}=B\boldsymbol{j}-D$,
         initial stress $\boldsymbol{\sigma}_0$,
         expansion point $s_0$, reduction order $q$
\Ensure Projection matrix $V_r$, reduced matrices $A_h$, $C_h$, $\boldsymbol{f}_h$
\State Form $K_{s_0} \leftarrow s_0 C - A$ and compute its sparse LU factors
\State Define the propagation operator $\mathcal{M}(X) \leftarrow K_{s_0}^{-1} C X$
\State Form starting block
       $F_0 \leftarrow \bigl[K_{s_0}^{-1}C\boldsymbol{\sigma}_0,\;
                              K_{s_0}^{-1}\boldsymbol{f}\bigr]$
\State Compute thin QR: $F_0 = V_1 R_0$; initialize $V \leftarrow V_1$
\vspace{0.3em}
\For{$j = 1$ to $q-1$}
    \State $W \leftarrow \mathcal{M}(V_j)$
    \For{$i = 1$ to $j$}
        \State $H_{ij} \leftarrow V_i^{\top} W$
        \State $W \leftarrow W - V_i H_{ij}$
    \EndFor
    \State Compute thin QR: $W = V_{j+1} H_{j+1,j}$
    \If{$\|H_{j+1,j}\|_F < \varepsilon$} \textbf{break} \EndIf
    \State Append $V_{j+1}$ to $V$
\EndFor
\vspace{0.3em}
\State $A_h \leftarrow V^{\top} A V$,\quad
       $C_h \leftarrow V^{\top} C V$,\quad
       $\boldsymbol{f}_h \leftarrow V^{\top}\boldsymbol{f}$
\State \Return $(V,\, A_h,\, C_h,\, \boldsymbol{f}_h)$
\end{algorithmic}
\end{algorithm}

\subsubsection{Reduced model and time integration}

Using the final basis $V_r$, the reduced matrices are
\begin{equation}
  \small
    A_h = V_r^{\top} A V_r, \qquad
    C_h = V_r^{\top} C V_r, \qquad
    \boldsymbol{f}_h = V_r^{\top}\boldsymbol{f},
    \label{eq:rak_reduced}
\end{equation}
and the reduced descriptor system
\[
    C_h\dot{\hat{\boldsymbol{\sigma}}}(t)
    = A_h\hat{\boldsymbol{\sigma}}(t) + \boldsymbol{f}_h,
    \qquad
    \hat{\boldsymbol{\sigma}}(0) = V_r^{\top}\boldsymbol{\sigma}_0,
\]
is integrated with backward Euler.  The full-space stress field is recovered
by the lift $\boldsymbol{\sigma}(t) \approx V_r\hat{\boldsymbol{\sigma}}(t)$.

\subsubsection{Shift time selection}

Rather than specifying $s_0$ directly, \textit{EMSpice~3} exposes the
\emph{shift time} $t_s = 1/s_0$, which has the physical interpretation of the
time-constant region of greatest interest.  An initial estimate is derived
from the analytical solution of the 1-D diffusion equation with blocking
boundary conditions at both ends.  The slowest-decaying eigenmode has
eigenvalue $\lambda_1 = (\pi/L)^2$, giving the diffusive time constant
\begin{equation}
  \small
    \tau = \frac{L^2}{\pi^2 \kappa},
    \label{eq:rak_shift_time}
\end{equation}
where $L$ is the wire length and $\kappa$ is the stress diffusivity. In the
\textit{EMSpice~3} flow, we set $L = L_{\text{max}}$ (maximum
path length in the tree), reflecting the longer-range void growth along the
critical path.  The shift time is set as $t_s = \eta\tau$, where $\eta$
is a user-configurable scaling factor.  In practice, setting $\eta = 1$
achieves less than 1\% error in nucleation time at reduction
order $q = 4$--$6$~\cite{LuTan2026RationalKrylovEM}.

\subsubsection{Implementation improvements}
Reduction techniques can be very effective for large trees, but they are not
uniformly beneficial across all extracted power-grid trees. In practical PDN
netlists, many trees are short, nearly straight segments with only one or two
branches. These trees are typically very cheap to solve directly: empirical
profiling in our implementation shows sub-millisecond runtime for tiny trees
and only about 1--3\,ms for small-to-medium trees. At that scale, the
overhead of the reduction process may exceed the numerical solve cost. As a
result, we propose three optimizations in \textit{EMSpice~3} to accelerate the
basic algorithm:

\textbf{Basis caching.}  The Arnoldi basis $V_r$ and reduced matrices
$(A_h, C_h, \boldsymbol{f}_h)$ are determined by the per-tree descriptor system,
the shifted starting block, and the shift $s_0$.
\textit{EMSpice~3} caches the basis in an LRU (least recently used) dictionary, where each entry is keyed by a fingerprint of the tree matrices together with $s_0$. Here, the fingerprint is a short, fixed-length hash value computed from the numerical data of the matrices and shift parameter; it uniquely identifies the input data so that cached results can be efficiently retrieved or invalidated if the inputs change. The expensive sparse
LU factorization and Arnoldi iterations run only once per tree; subsequent
outer steps retrieve the cached basis at negligible cost.

\textbf{Loop-free time stepping for uniform grids.}  When the inner time
grid is uniformly spaced with step $\Delta t$, the backward Euler recurrence
$\hat{\boldsymbol{\sigma}}_k = F\hat{\boldsymbol{\sigma}}_{k-1} + \boldsymbol{g}$
(where
$F = (C_h - \Delta t\,A_h)^{-1} C_h$ and
$\boldsymbol{g} = \Delta t\,(C_h - \Delta t\,A_h)^{-1}\boldsymbol{f}_h$)
is solved without a Python loop. $F$ is eigendecomposed once as
$F = E\,\mathrm{diag}(\lambda)\,E^{\top}$, and all time steps
are computed simultaneously via vectorized power-series operations:
\[
\hat{\boldsymbol{\sigma}}_k
= E\!\left(\lambda^k \odot \boldsymbol{c}_0
         + \textstyle\sum_{j<k}\lambda^j \odot \boldsymbol{c}_g\right),
\]
where $\boldsymbol{c}_0 = E^{\top}\hat{\boldsymbol{\sigma}}_0$ and
$\boldsymbol{c}_g = E^{\top}\boldsymbol{g}$.  The inner cumulative sum is
computed via \texttt{np.cumsum}, making the entire trajectory a
sequence of array operations with no per-step overhead.  For non-uniform
grids, the LU factorization of $(C_h - \Delta t\,A_h)$ is cached per unique
$\Delta t$ and reused across steps.

\textbf{Batch lift.}  The $n_t \times r$ reduced-space trajectory is lifted
to full space in a single BLAS \textsc{dgemm}:
$\text{series} = \hat{Z}\,V_r^{\top}$,
replacing $n_t$ individual $O(nr)$ matrix-vector products with one
cache-friendly $O(n_t n r)$ matrix multiply.

\subsubsection{Fallback to backward Euler and dynamic tree selection}

Not all trees benefit from Krylov reduction. \textit{EMSpice~3} therefore
applies a \emph{dynamic size threshold} before invoking the solver.  In the
implementation, this threshold depends not only on the state dimension $n$ and
reduction order $q$, but also on whether a Krylov basis is already cached and
on the number of inner timesteps to be advanced. On a cold solve, the basis
construction and shifted sparse factorization are worthwhile only when $n$ is
sufficiently larger than $q$; specifically, the Krylov path is selected when
\begin{equation}
  \small
  n \geq \max(3q,\;30),
  \label{eq:krylov_threshold}
\end{equation}
ensuring the full-order system is large enough relative to the reduction order
that the setup cost is worthwhile. When the cache is warm, the threshold can be
reduced substantially because the expensive setup is skipped. This policy is
important in practice: many trees are small enough that direct backward Euler
already costs only a few milliseconds, whereas the minority of larger trees
benefit from repeated reduced-order solves across outer aging iterations.
Small trees therefore fall through to the lightweight full-order path with
negligible cost, while the expensive tail of the tree population is
accelerated. The accelerated path can also be disabled globally, in which case
full backward Euler is used throughout.

Even for trees that pass the size threshold, the solver may encounter
numerical difficulty (e.g., rank deficiency in the Arnoldi basis or
ill-conditioned shift).  When this occurs, the tree is transparently
re-solved with full backward Euler.  Krylov reduction is therefore a
\emph{best-effort} performance optimization: it accelerates the common case
without risk to correctness.

\subsection{Monte Carlo statistical lifetime analysis}
\label{subsec:mc_framework}

Beyond deterministic EM/TM simulation, the framework also supports statistical
reliability analysis through Monte Carlo sampling. In this mode, the full
analysis pipeline is re-executed for each sample, including immortality
screening, transient stress evolution, resistance update, and IR-drop-based
TTF evaluation. The output is therefore not a single lifetime estimate, but a
distribution of TTF values together with the corresponding counts of censored
and failed samples.

In the current implementation used in this paper, we perturb the 
EM diffusivity~$\kappa(x)$ and the critical stress with a prescribed CoV, then run 100 independent samples for each design.
The numerical results in Section~\ref{sec:observations} use a 20\% CoV for both parameters,
with each design simulated under its 353\,K Joule-heating thermal condition
(the first input thermal-map case for each respective design),
and failure defined by the same 10\% IR-drop threshold as in the deterministic
runs. This setup directly
measures how process variation changes the predicted TTF distribution while
keeping the electrical and thermal workloads fixed.

The sampling framework itself is more general than the two-parameter study
reported here. Any model input that can be mapped into the per-sample netlist,
material file, or solver parameter set can in principle be treated as a random
variable. Examples include resistivity, effective diffusivity, critical stress,
activation-energy-related parameters, geometric dimensions, current loading,
and even external temperature-map perturbations. From the solver viewpoint,
these are simply alternate sampled inputs for the same coupled EM/TM/IR-drop
flow, so no algorithmic change is required once the sampled values are stamped
into the corresponding data structures before each run.

This separation between the sampling layer and the physics solver is important
for extensibility. It allows the present paper to focus on variation in
$\kappa(x)$ and critical stress, which already reveals the strong
design-dependence reported in the Monte Carlo results, while keeping the door
open to richer statistical reliability studies that incorporate additional
process, thermal, electrical, or workload-induced uncertainty sources.

\section{Numerical results and case studies}
\label{sec:observations}
In this section, we present numerical case studies and comparative evaluations
performed using the proposed framework as implemented in \textit{EMSpice~3}.
All simulations were run on a high-performance Linux server equipped with dual
3.3\,GHz Intel Xeon processors and 316\,GB of memory.

\subsection{Case Study I: RISC-V Core}
\begin{figure}[!htb]
  \centering
  \begin{subfigure}[b]{0.47\columnwidth}
    \centering
    \includegraphics[width=\linewidth,height=1.25in]{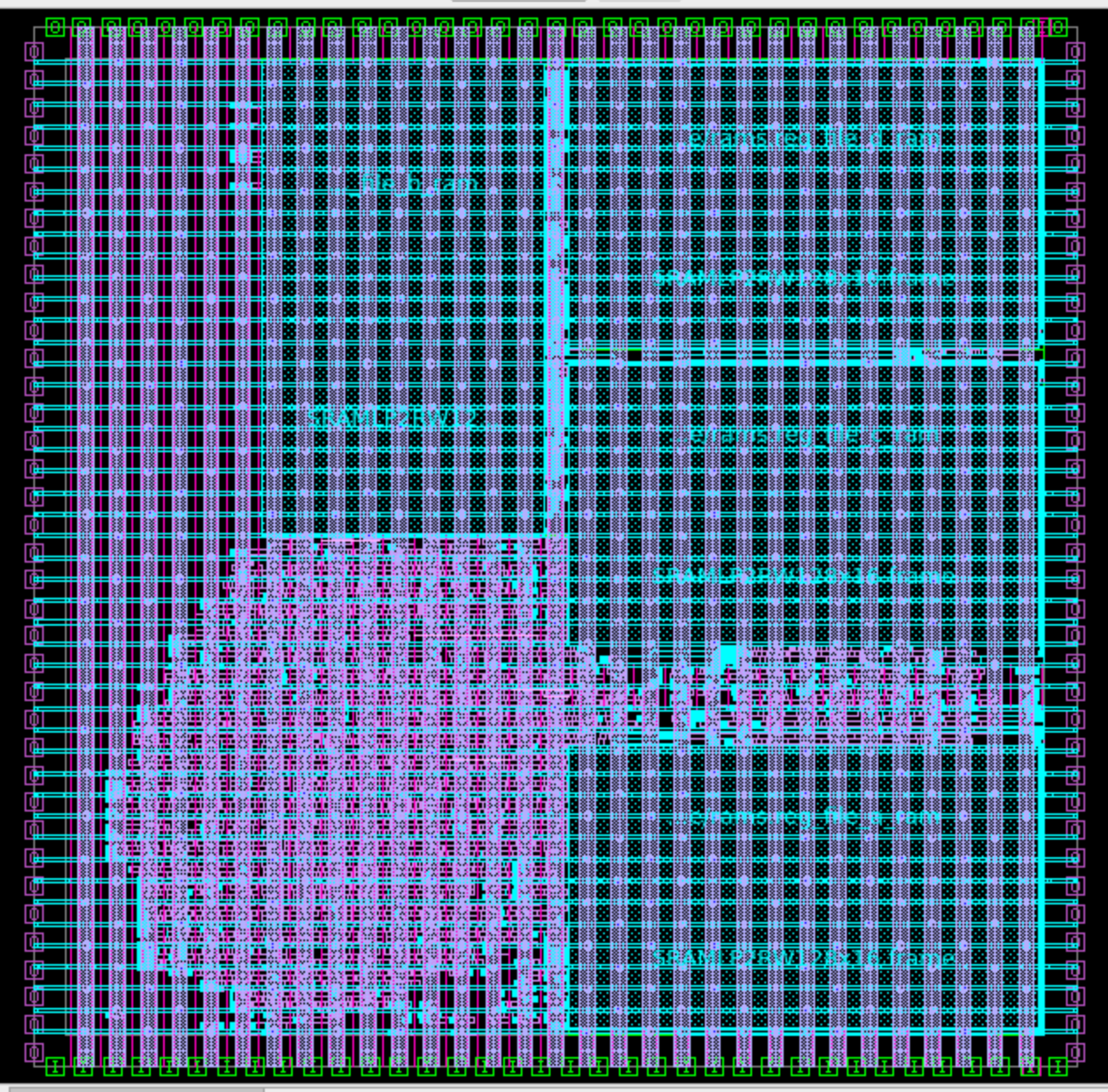}
    \caption{\small RISC-V core: finished physical layout}
    \label{fig:risc_core_layout}
  \end{subfigure}%
  \begin{subfigure}[b]{0.49\columnwidth}
    \centering
    \includegraphics[width=\linewidth,height=1.25in]{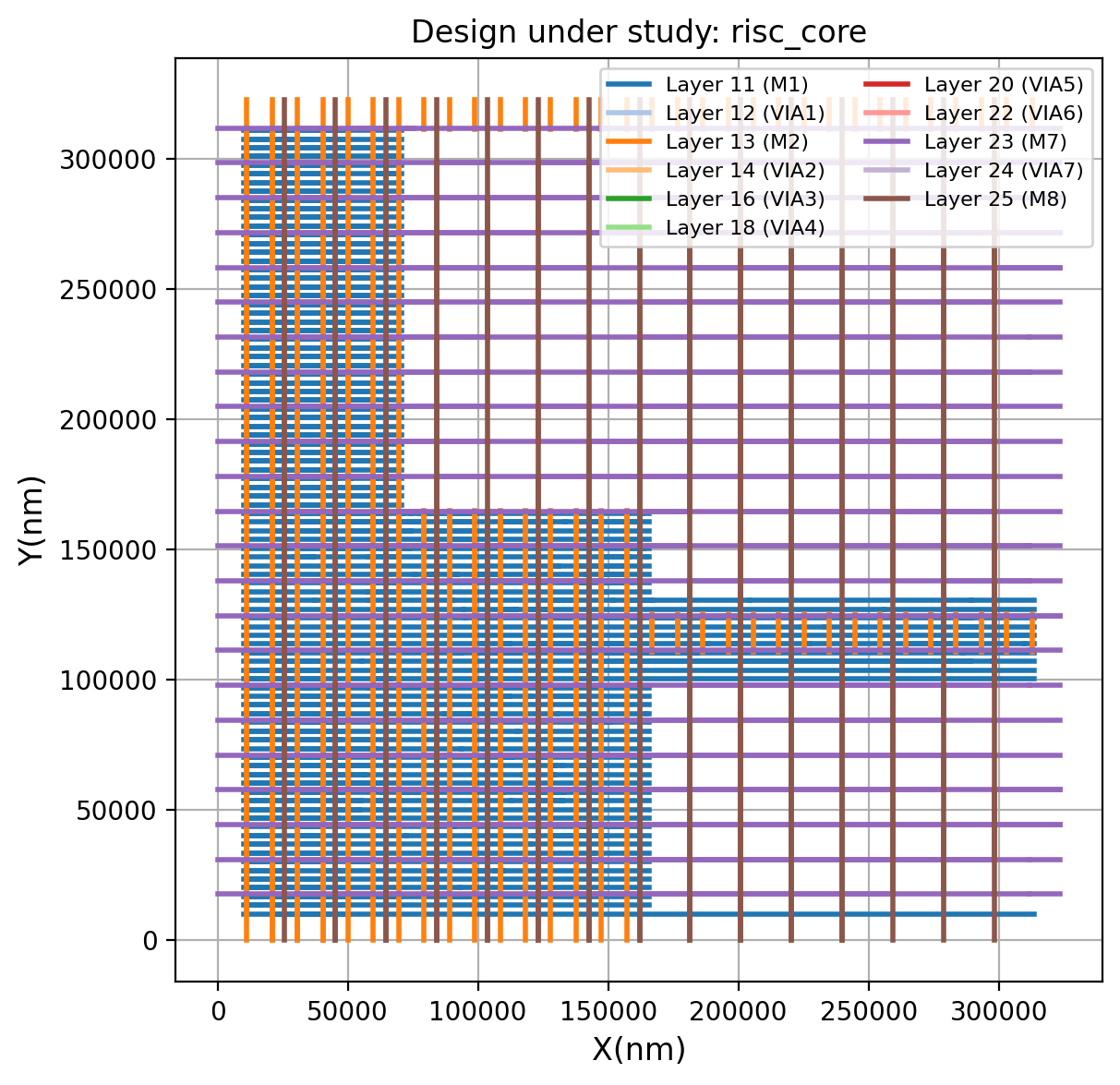}
    \caption{\small VDD Power grid structure}
    \label{fig:risc_core_design}
  \end{subfigure}
  \caption{\small RISC-V core: finished physical layout (left) and extracted
  power grid structure showing four metal layers (M1, M2, M7, M8) with VIA
  connections (right).}
  \label{fig:risc_core_overview}
\end{figure}

Our first case study focuses on the power grid of a RISC-V processor core,
herein referred to as ``RISC''. This design was synthesized and placed-and-routed
using the SAED32 (Synopsys 32/28nm Generic) library~\cite{SynopsysGenericLibrary}. 
The design comprises
9 standard-cell blocks across 11 routing layers, with a total chip area of
104,385.65\,\textmu m$^2$ ($\approx$0.104\,mm$^2$). The RISC power grid spans four
metal layers (M1, M2, M7, and M8, interconnected by VIA layers) with a total
of 186 trees, 115 maximum branches per tree, and up to 5750 nodes per tree.
The supply voltage is 0.95\,V.
The finished layout and power grid structure are shown in
Fig.~\ref{fig:risc_core_overview}.
All simulations use a base temperature of 353\,K (80\,$^\circ$C) and a total
simulation horizon of $5\times10^7$\,s ($\approx$1.6\,years), with ten outer
timesteps of $5\times10^6$\,s each.

We run two scenarios that differ only in their spatial temperature
distribution, both with the same 353\,K average temperature.

\begin{figure}[!htb]
  \centering
  \begin{subfigure}[b]{0.48\columnwidth}
    \centering
    \includegraphics[width=\linewidth]{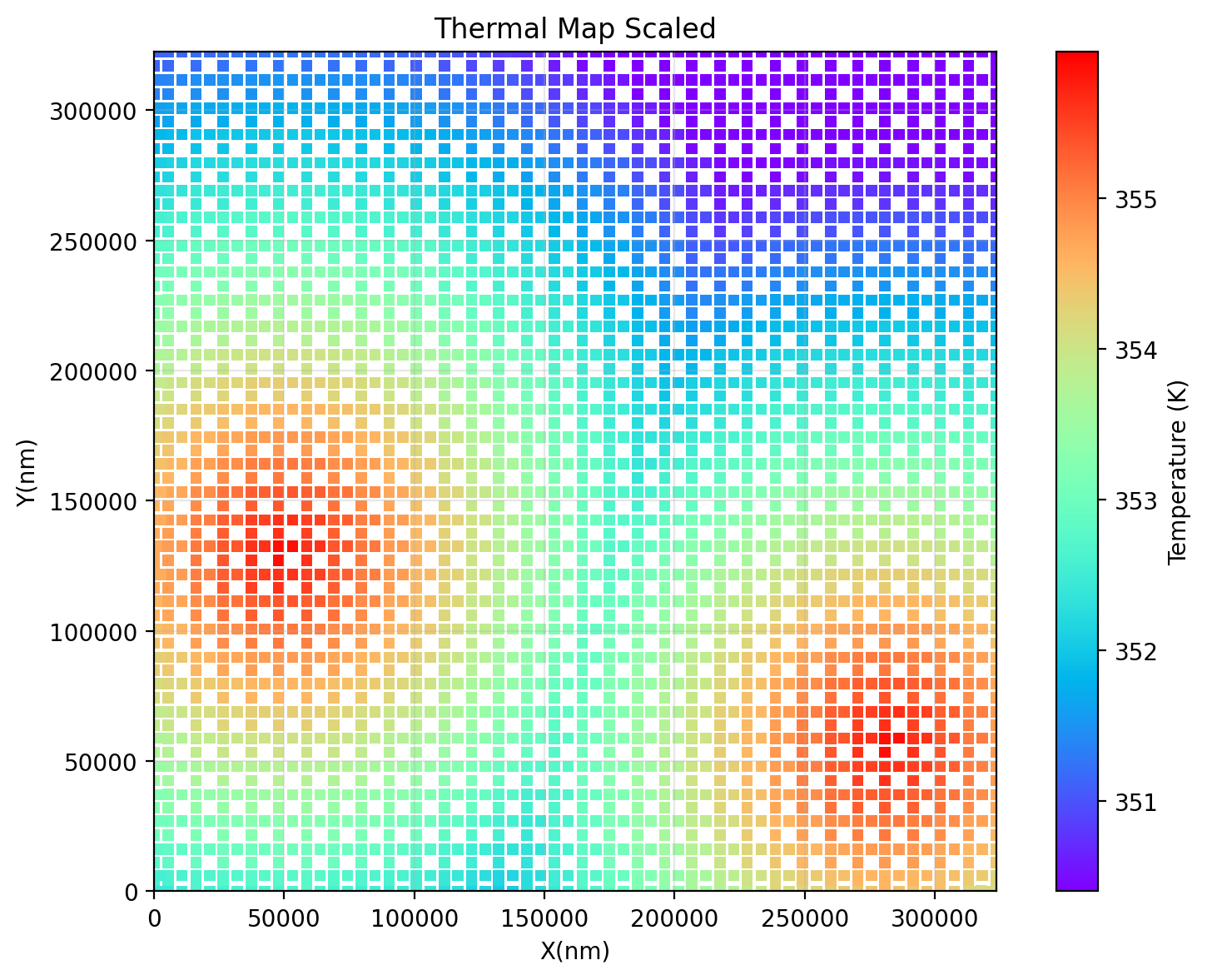}
    \caption{\small Externally given baseline temperature map (Case 1):
    localized lower-left hotspot, peak $\approx$355\,K, remainder $\approx$351\,K}
    \label{fig:risc_baseline_tmap}
  \end{subfigure}%
  \hfill
  \begin{subfigure}[b]{0.48\columnwidth}
    \centering
    \includegraphics[width=\linewidth]{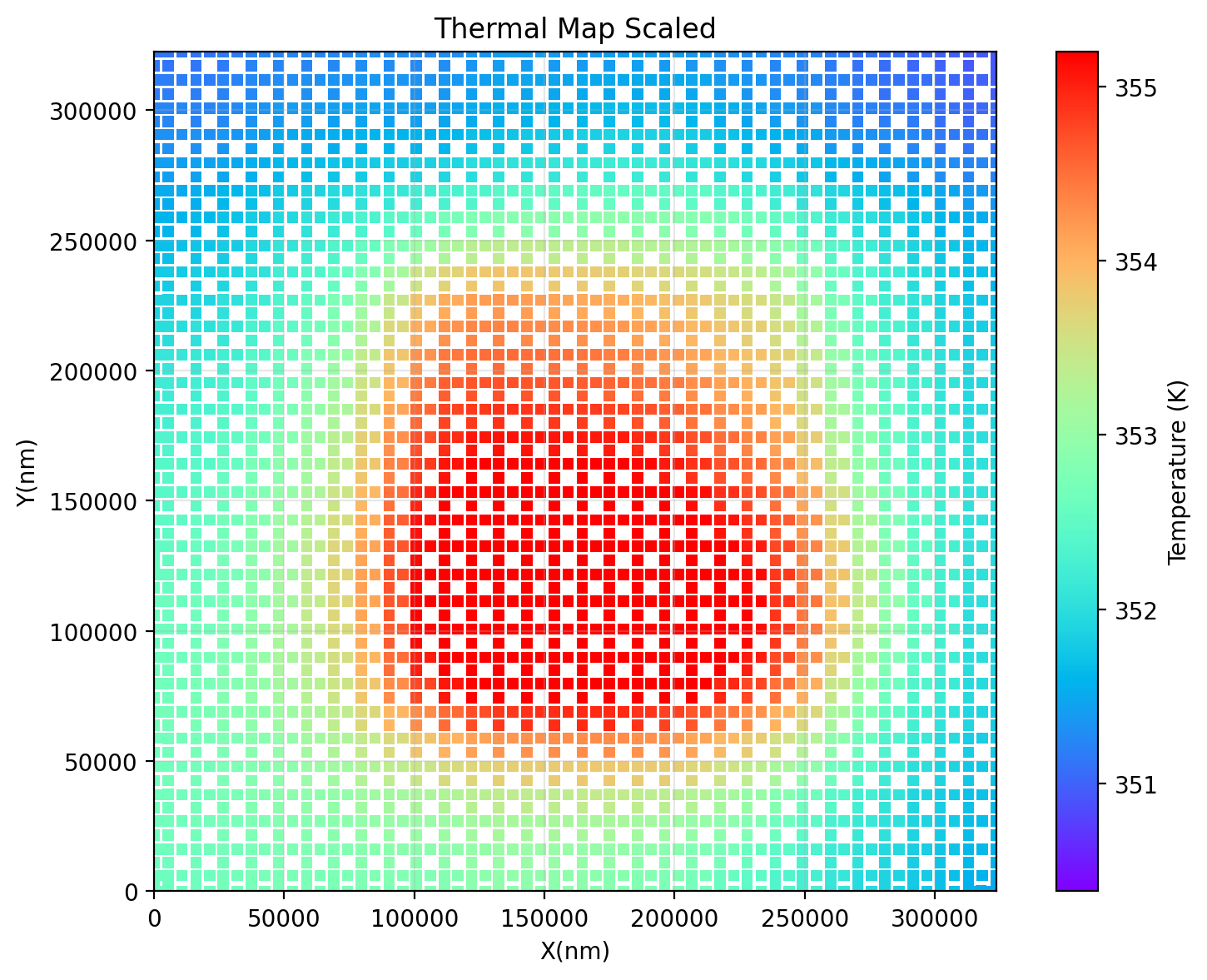}
    \caption{\small Externally given Qualcomm SM6225 measured temperature map
    (Case 2): broad central hotspot, peak $\approx$355\,K, same 353\,K average}
    \label{fig:risc_qualcomm_tmap}
  \end{subfigure}
  \caption{\small Externally given spatial temperature maps provided as inputs
  to \textit{EMSpice~3} for the two RISC-V core scenarios. Both maps share a
  353\,K die-average temperature but differ in spatial structure: the baseline
  map has a compact lower-left hotspot; the Qualcomm measured map spreads heat
  broadly across the central die area.}
  \label{fig:risc_tmap}
\end{figure}

\begin{figure}[!htb]
  \centering
  \begin{subfigure}[b]{0.48\columnwidth}
    \centering
    \includegraphics[width=\linewidth]{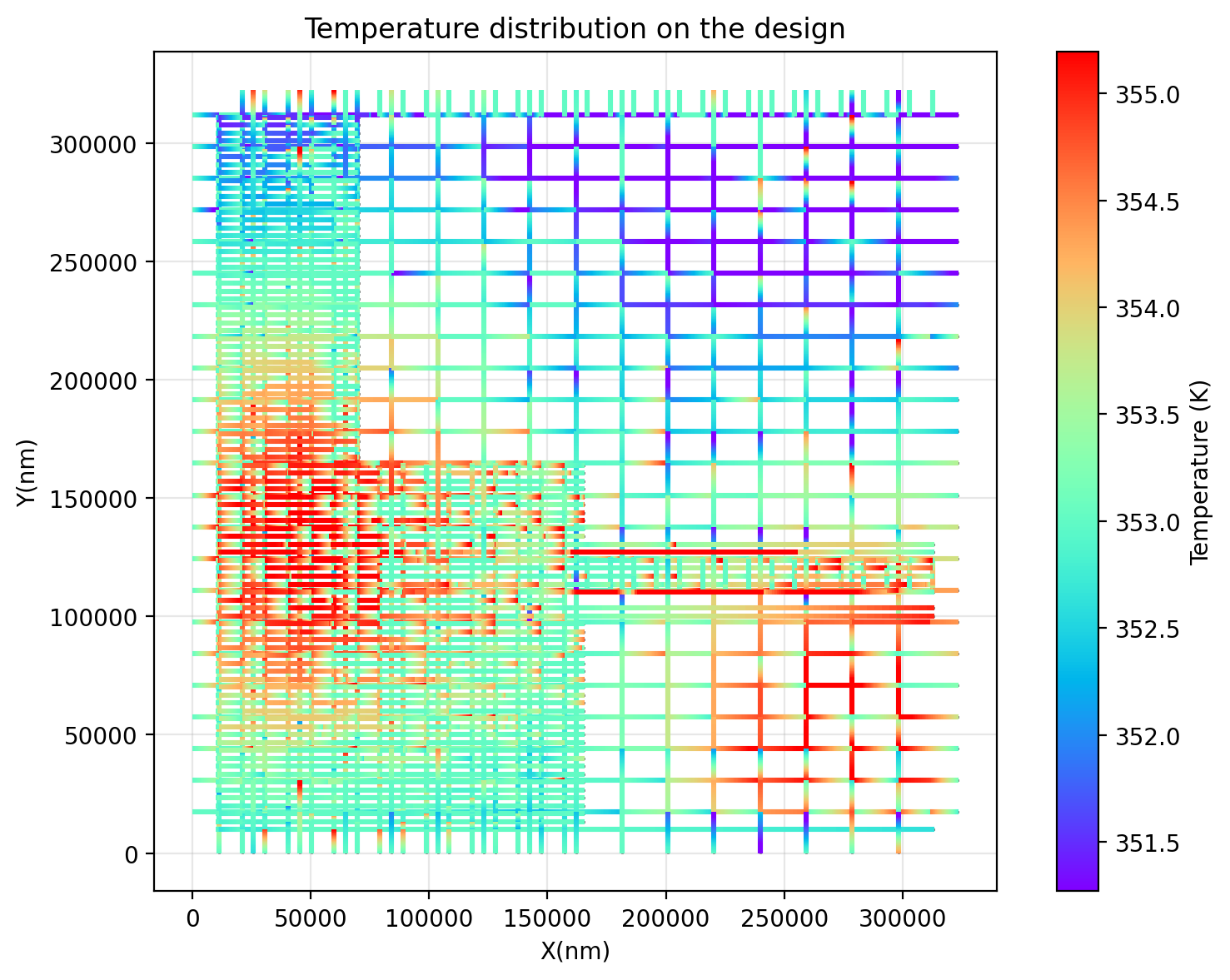}
    \caption{\small Per-node Joule-heating temperature distribution (Case 1):
    intense heating along high-current wire segments in the lower-left;
    external baseline hotspot and Joule gradient reinforce each other}
    \label{fig:risc_baseline_jht}
  \end{subfigure}%
  \hfill
  \begin{subfigure}[b]{0.48\columnwidth}
    \centering
    \includegraphics[width=\linewidth]{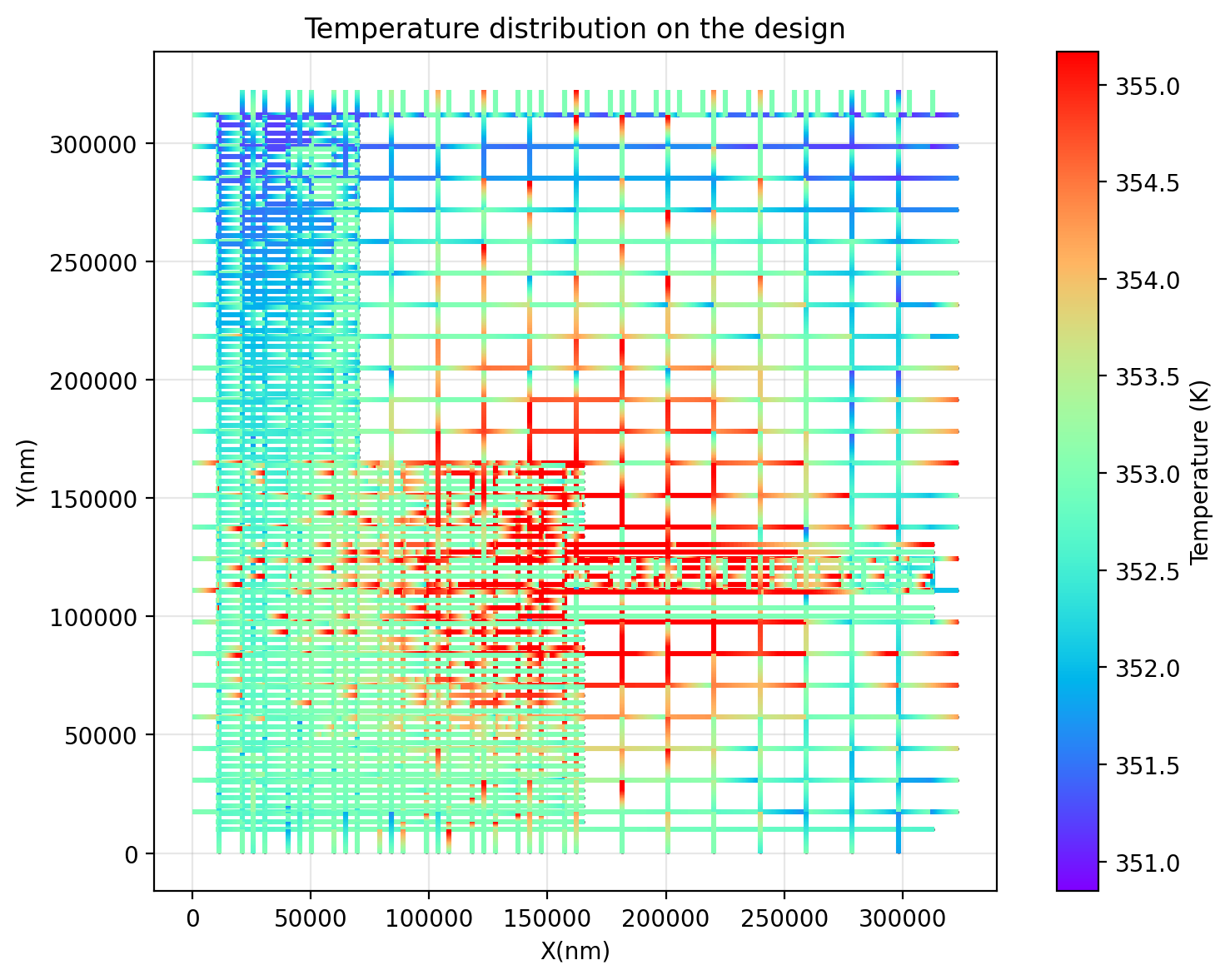}
    \caption{\small Per-node Joule-heating temperature distribution (Case 2):
    similar wire-level pattern as Case~1 but modulated by the broad Qualcomm
    ambient, which raises base temperature centrally rather than over the
    highest-current wires}
    \label{fig:risc_qualcomm_jht}
  \end{subfigure}
  \caption{\small Per-node Joule-heating temperature distributions on the
  RISC-V power grid computed by \textit{EMSpice~3} for the two thermal
  scenarios. Individual wire segments are resolved, showing the spatial
  correlation between current density and local temperature. In Case~1 the
  external baseline hotspot directly overlaps the most current-stressed region;
  in Case~2 the Qualcomm ambient heat does not preferentially load the
  highest-current power-delivery wires.}
  \label{fig:risc_jht_tmap}
\end{figure}

\begin{figure*}[!t]
  \centering
  \begin{subfigure}[b]{0.34\textwidth}
    \centering
    \includegraphics[width=\linewidth]{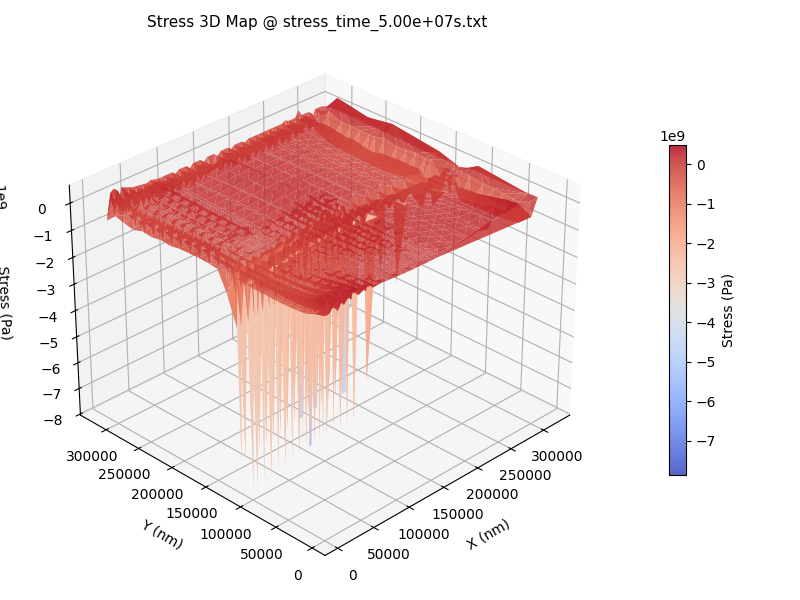}
    \caption{\small 3D Stress map and nucleation sites (Case 1): 18 mortal trees
    at $5\times10^7$\,s; final stress field predominantly compressive}
    \label{fig:risc_baseline_stress}
  \end{subfigure}%
  \hspace{0.04\textwidth}
  \begin{subfigure}[b]{0.34\textwidth}
    \centering
    \includegraphics[width=\linewidth]{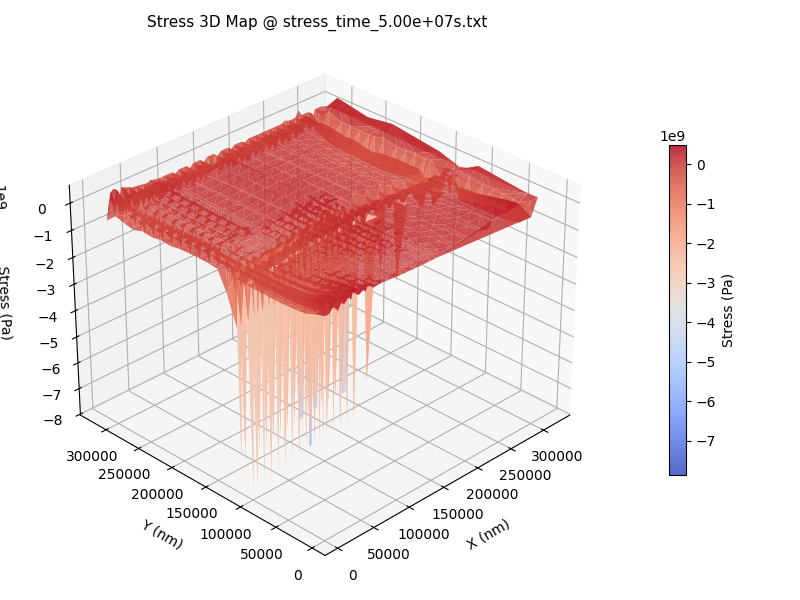}
    \caption{\small 3D Stress map and nucleation sites (Case 2): 16 mortal trees
    at $5\times10^7$\,s; final stress field predominantly compressive}
    \label{fig:risc_qualcomm_stress}
  \end{subfigure}

  \begin{subfigure}[b]{0.34\textwidth}
    \centering
    \includegraphics[width=\linewidth]{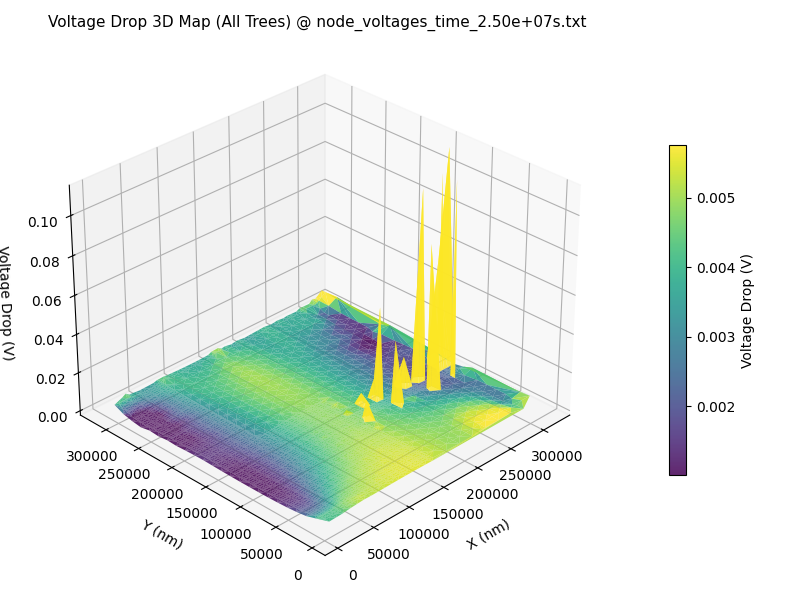}
    \caption{\small 3D IR-drop map at $t=2.50\times10^7$\,s (Case 1):
    max IR-drop 12.86\%}
    \label{fig:risc_baseline_vdrop}
  \end{subfigure}%
  \hspace{0.04\textwidth}
  \begin{subfigure}[b]{0.34\textwidth}
    \centering
    \includegraphics[width=\linewidth]{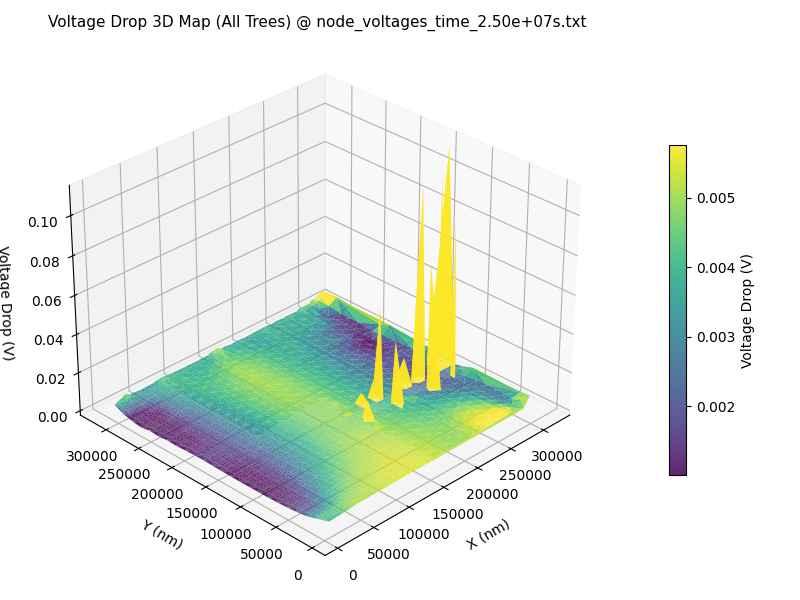}
    \caption{\small IR-drop distribution for the Qualcomm-map case (Case 2)
    during late-stage degradation}
    \label{fig:risc_qualcomm_vdrop}
  \end{subfigure}

  \caption{\small Stress distributions and IR-drop maps for the RISC-V core
  under the two thermal scenarios (external input maps: Fig.~\ref{fig:risc_tmap};
  Joule-heating distributions: Fig.~\ref{fig:risc_jht_tmap})}
  \label{fig:risc_all}
\end{figure*}

\textbf{Case 1 — Baseline external temperature map.}
The first scenario uses a baseline spatial temperature distribution provided
externally as input to \textit{EMSpice~3}. As shown in
Fig.~\ref{fig:risc_baseline_tmap}, this map features a localized hotspot
concentrated in the lower-left region of the die, peaking at approximately
355\,K while the rest of the chip stays near 351\,K. The per-node
Joule-heating temperature distribution computed by \textit{EMSpice~3} under
this external map is shown in Fig.~\ref{fig:risc_jht_tmap}(a).
The steady-state immortality screening identifies 7 trees as immortal before the
transient simulation begins.
During the subsequent transient FDTD solve, 18 trees nucleate voids by the end
of the simulation. The resulting stress distribution and nucleation sites are
illustrated in Fig.~\ref{fig:risc_baseline_stress}.
The initial maximum IR-drop is 6.19\% of $V_{src}$; it rises to 29.64\% by
the end of the simulation. The IR-drop 10\% failure threshold is crossed at
approximately $2.47\times10^7$\,s ($\approx$9.4 months);
Fig.~\ref{fig:risc_baseline_vdrop} shows the voltage drop map at $t=2.50\times10^7$\,s. 

\begin{figure}[!htb]
  \centering
  \includegraphics[width=0.80\linewidth]{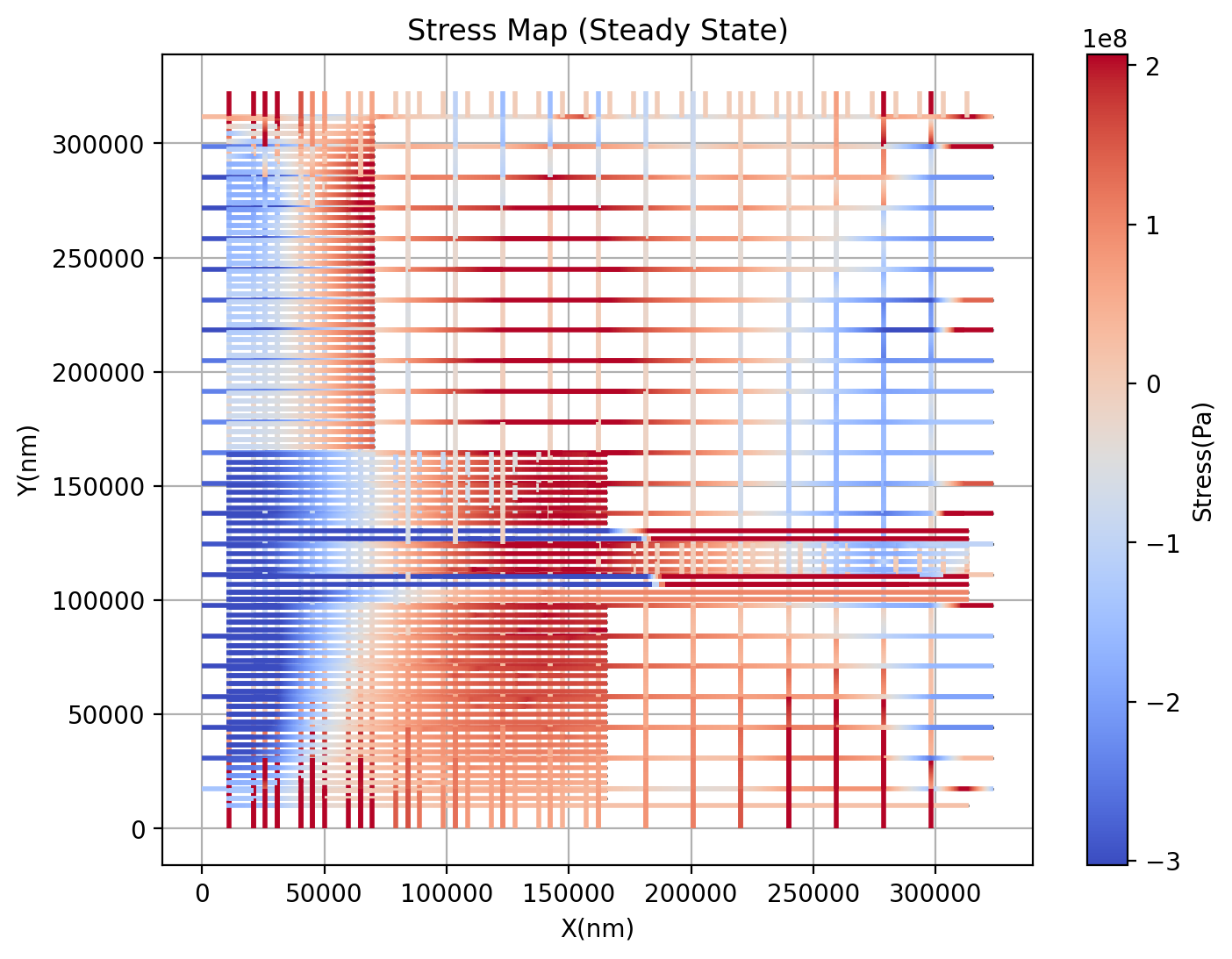}
  \caption{\small Steady-state stress distribution for the RISC-V core (Case~1,
  baseline external temperature map at 353\,K average) computed without void
  nucleation. Stress accumulates unbounded throughout the simulation, revealing
  the segments most susceptible to EM degradation prior to any saturation effect.}
  \label{fig:risc_stress_steady}
\end{figure}

Fig.~\ref{fig:risc_stress_steady} shows the steady-state stress distribution
for Case~1 computed \emph{without} void nucleation. Without the saturation
mechanism, stress accumulates continuously and peak tensile values in the
lower-left hotspot far exceed those in Fig.~\ref{fig:risc_baseline_stress},
where void nucleation clamps stress near the saturation threshold once flux
is interrupted. This comparison highlights the importance of accurately
modeling nucleation in coupled EM/TM simulations.

\textbf{Case 2 — Qualcomm measured external temperature map.}
The second scenario uses a real spatial temperature distribution measured from
a Qualcomm SM6225 Snapdragon 680 4G 8-core CPU chip, provided as an external
input at the same 353\,K average temperature~\cite{ucr_thermal_map_dataset}. As
shown in Fig.~\ref{fig:risc_qualcomm_tmap}, this map features a broad central
hotspot covering a large area of the die, rather than the localized corner
hotspot of Case~1. The per-node Joule-heating temperature distribution computed
by \textit{EMSpice~3} under this external map is shown in
Fig.~\ref{fig:risc_jht_tmap}(b).
Despite the wider heated region, the transient simulation identifies only 16
mortal trees (Fig.~\ref{fig:risc_qualcomm_stress}), two fewer than Case~1.
In the updated thermal sweep summarized in Tab.~\ref{tab:risc_thermal}, the
Qualcomm 353\,K case does not reach the 10\% IR-drop threshold within the
simulation horizon. This indicates a less severe degradation trajectory than
Case~1 under the same average temperature, as illustrated by the IR-drop map in
Fig.~\ref{fig:risc_qualcomm_vdrop}.

Fig.~\ref{fig:risc_tmap} shows the two input spatial temperature maps.
Despite sharing the same 353\,K average, they differ fundamentally: the
baseline map (Fig.~\ref{fig:risc_baseline_tmap}) has a compact lower-left
hotspot peaking at $\approx$355\,K, while the Qualcomm SM6225 measured map
(Fig.~\ref{fig:risc_qualcomm_tmap}) distributes heat over a broad central
region. Fig.~\ref{fig:risc_jht_tmap} shows the resulting per-node
Joule-heating distributions. In Case~1, the external baseline hotspot
directly overlaps the highest-current wires in the lower-left, so the two
gradients reinforce each other and compound the EM risk for those trees.
In Case~2, the Qualcomm ambient peak is not aligned with the
highest-current wires, so the combined thermal load on the most stressed
segments is lower—explaining the fewer mortal trees (16 vs.\ 18) and
later TTF despite the identical average temperature.

The spatial pattern of the temperature distribution, rather than its average
value, governs EM reliability: the baseline external map concentrates its
hotspot over the highest-current trees near the power delivery boundary,
aligning the thermal and electrical stress peaks and producing the most severe
Joule-heating on those wires (Fig.~\ref{fig:risc_baseline_jht}), while the
Qualcomm map spreads ambient heat over a broader central region that does not
preferentially overlap the most stressed wires, leading to fewer mortal trees
and no 10\% IR-drop threshold crossing for the 353\,K and 373\,K Qualcomm
profiles within the simulated horizon.

Tab.~\ref{tab:risc_thermal} extends the comparison to seven thermal profiles.
The mortal tree count varies from 7 (uniform 353\,K) to 18 (baseline
Joule-heating-derived external map at 353\,K), confirming that the spatial
structure—not the mean temperature—controls which trees fail. Among the 353\,K
profiles, the uniform case fails at $1.32\times10^7$\,s, the baseline external
map fails at $2.47\times10^7$\,s, and the Qualcomm map does not reach the
10\% IR-drop threshold within the simulation horizon. Although the uniform
profile yields the fewest mortal trees, the spatially varying maps activate
many more mortal trees, and the Qualcomm profile remains below the failure
criterion despite having 16 mortal trees. Notably, the two 353\,K nonuniform
profiles have similar maximum temperatures, 387.11\,K for the baseline map and
387.57\,K for the Qualcomm map, yet only the baseline map reaches the 10\%
threshold. At 373\,K average temperature, the two maxima again remain close,
407.26\,K and 407.72\,K, while the Qualcomm case still does not reach the
failure threshold. This indicates that hotspot alignment with critical current
paths can dominate over both tree count and scalar maximum temperature.

Tab.~\ref{tab:risc_thermal} also exposes a more specific mechanism: the
TTF of the power grid is governed not only by how many trees become mortal, but also by
where the first void is placed inside a critical tree. In the uniform 353\,K
case, tree R11-135 nucleates at cathode node 0. The resulting void degradation
first updates branch R11-135-120 and later branch R11-135-100, driving the
final maximum IR-drop to 17.745\% and producing a TTF of
$1.324214\times10^7$\,s. Under the Qualcomm 353\,K map, the same tree
R11-135 nucleates instead at cathode node 9, so only branch R11-135-100 is
degraded; the final maximum IR-drop remains 9.874\% and the 10\% threshold is
not reached. The Qualcomm 373\,K map follows the same void-location pattern,
with a slightly higher final maximum IR-drop of 9.887\%, still below the
failure threshold. This comparison shows that the mapped temperature field
changes the coupled EM/TM stress solution, including temperature-dependent
diffusivity and thermomigration driving forces, so the peak-stress location
selected for nucleation can move within the same tree. Once the void location
is selected, the degraded resistor branch controls the network resistance
perturbation and therefore dominates the IR-drop TTF. This branch-selection
effect is a key reason why average temperature, maximum temperature, and mortal
tree count alone are insufficient for sign-off.

\begin{table}[!htb]
  \centering
  \caption{\small RISC-V core EM results under different thermal profiles
  (186 trees total). TTF of the power grid is the time at which the network-wide IR-drop exceeds
  the 10\% failure threshold.}
  \label{tab:risc_thermal}
  \footnotesize
  \setlength{\tabcolsep}{2pt}
  \begin{tabular}{c|cccc}
    \hline\hline
    Thermal Profile &
      \shortstack{Avg.\\Temp (K)} &
      \shortstack{Max\\Temp (K)} &
      \shortstack{Mortal\\Trees} &
      TTF (s) \\
    \hline\hline
    Uniform 353\,K                    & 353.00 & 353.00 &  7 & $1.32\times10^7$ \\
    Joule-heating (avg.\ 320\,K) & 320.62 & 345.93 & 14 & $3.01\times10^7$ \\
    Joule-heating (avg.\ 353\,K) & 353.40 & 387.11 & 18 & $2.47\times10^7$ \\
    Joule-heating (avg.\ 373\,K) & 373.55 & 407.26 & 18 & $1.45\times10^7$ \\
    Qualcomm map (avg.\ 309\,K)  & 309.63 & 336.90 & 12 & $1.87\times10^7$ \\
    Qualcomm map (avg.\ 353\,K)  & 353.23 & 387.57 & 16 & \shortstack{Not reach 10\%\\of $V_{dd}$} \\
    Qualcomm map (avg.\ 373\,K)  & 373.38 & 407.72 & 16 & \shortstack{Not reach 10\%\\of $V_{dd}$} \\
    \hline\hline
  \end{tabular}
\end{table}

\subsection{Case Study II: ARM Cortex-A Logic Core}
\begin{figure}[!htb]
  \centering
  \includegraphics[width=0.8\linewidth]{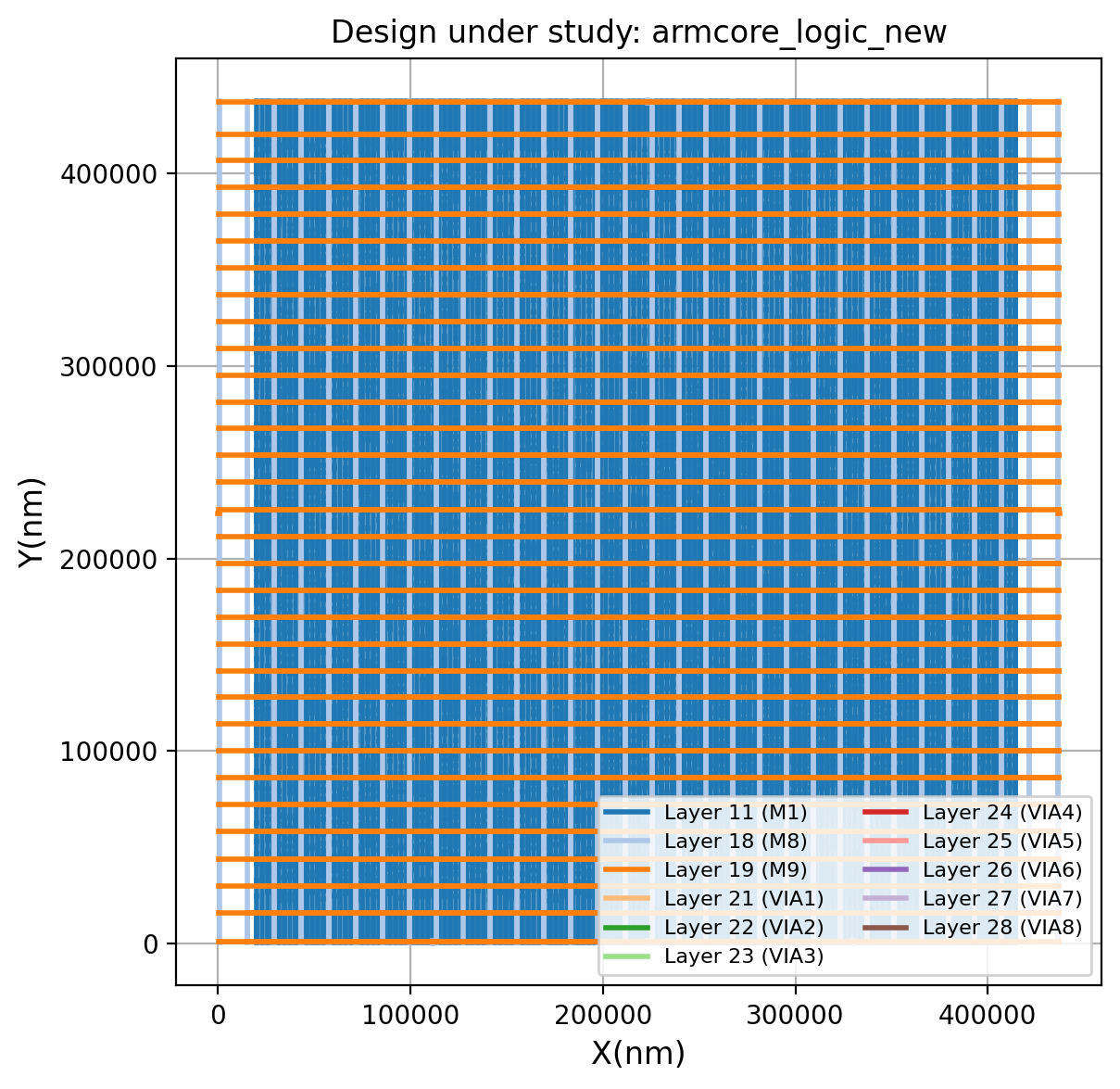}
  \caption{\small Power grid structure of the ARM Cortex-A logic core (three
  metal layers M1, M8, M9 shown together with VIA connections)}
  \label{fig:armcore_design}
\end{figure}

The second test case focuses on the power grid of an ARM Cortex-A logic core
design, herein referred to as ``ARMcore''. The design was synthesized and
placed-and-routed using the Synopsys Design Compiler with the Synopsys
32/28nm Generic Library~\cite{SynopsysGenericLibrary}. The ARMcore power
grid spans three metal layers (M1/L11, M8/L18, and M9/L19, connected by
VIA layers L21--L28) with a total of 208 trees, a maximum of 218 branches
per tree, and up to 10\,900 nodes per tree. The supply voltage is 1.2\,V.
The power grid structure spanning a $450\times450\,\mu$m die footprint is
shown in Fig.~\ref{fig:armcore_design}. All simulations use a base
temperature of 353\,K (80\,$^\circ$C) and a total simulation horizon of
$5\times10^7$\,s ($\approx$1.6\,years), with ten outer timesteps of
$5\times10^6$\,s each.

We again run two scenarios that differ only in their spatial temperature
distribution, both with the same 353\,K average temperature.

\begin{figure*}[!t]
  \centering
  \begin{subfigure}[b]{0.34\textwidth}
    \centering
    \includegraphics[width=\linewidth]{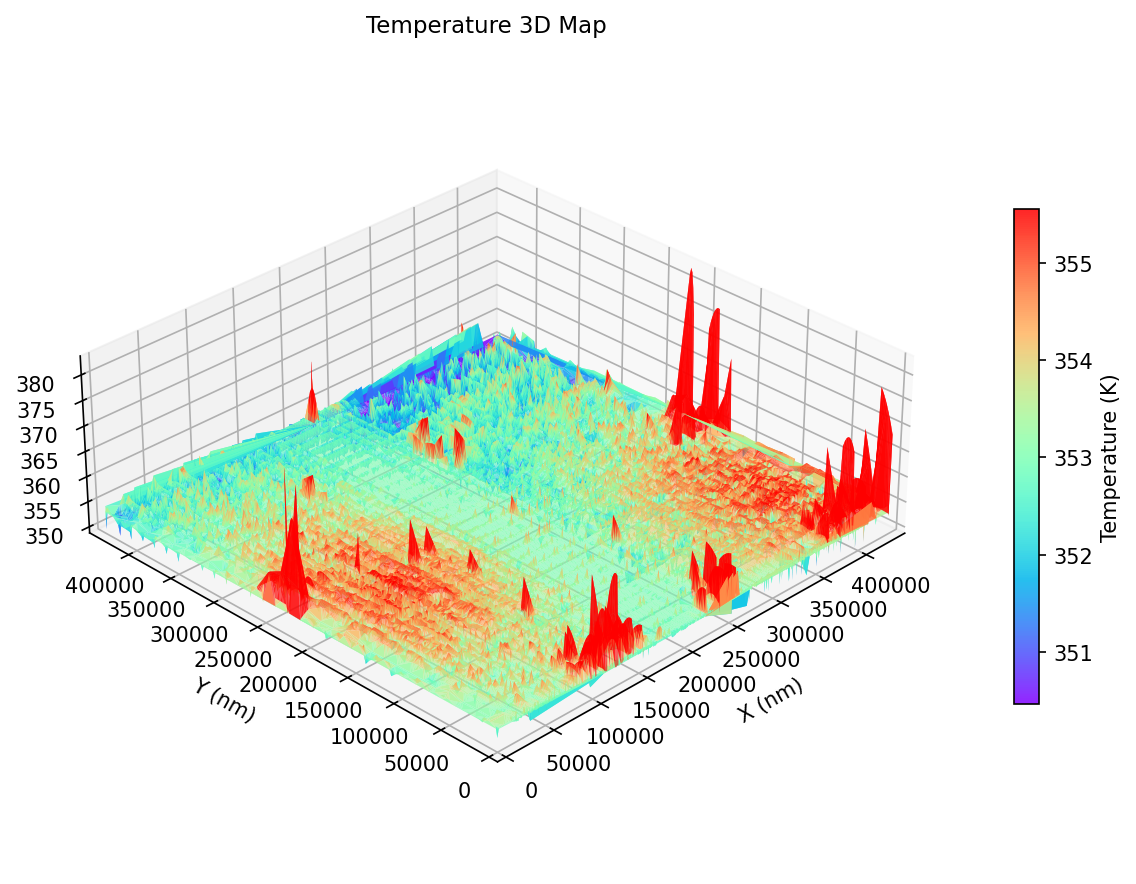}
    \caption{\small Baseline external 3D temperature distribution (Case 1):
    the displayed color scale highlights the bulk field near 353\,K, with
    warmer lower-left and lower-right regions and a cooler upper-right corner}
    \label{fig:armcore_baseline_tmap}
  \end{subfigure}%
  \hspace{0.04\textwidth}
  \begin{subfigure}[b]{0.34\textwidth}
    \centering
    \includegraphics[width=\linewidth]{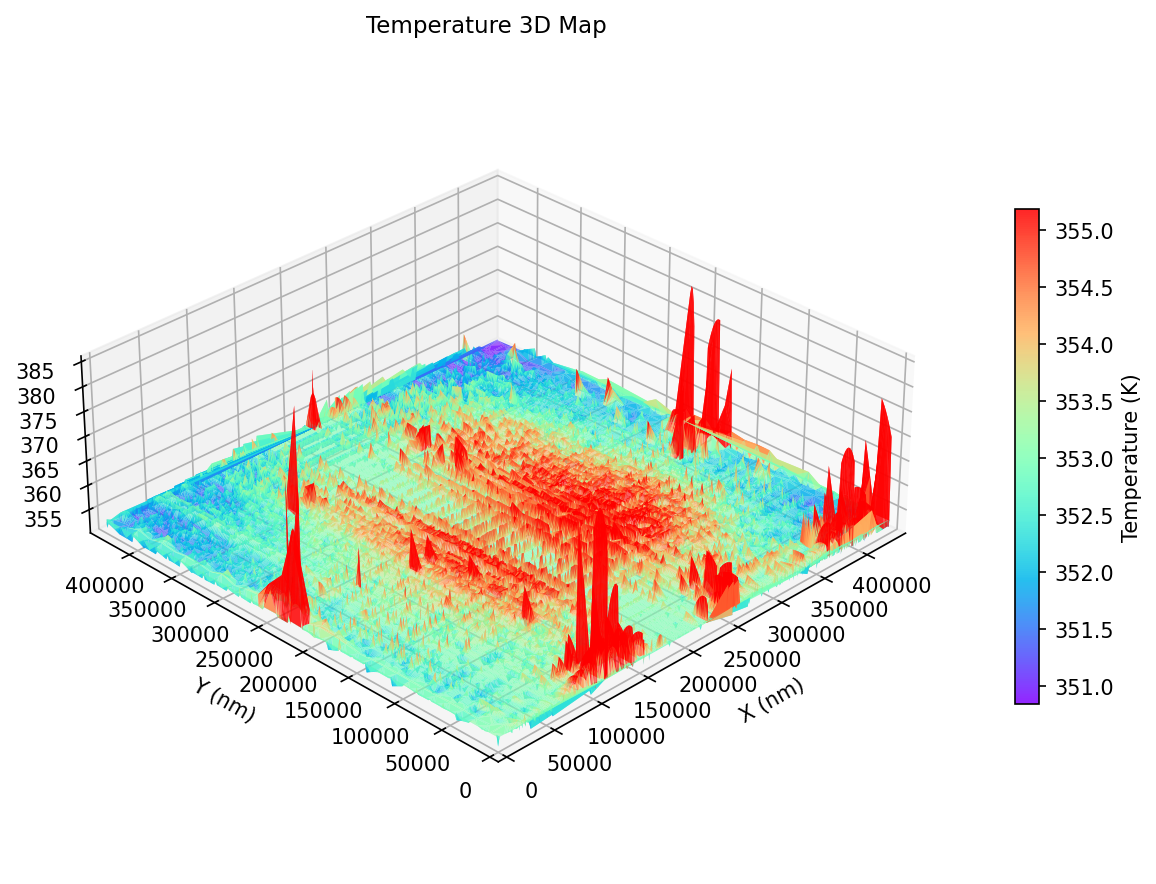}
    \caption{\small Qualcomm measured 3D temperature distribution (Case 2):
    the displayed color scale highlights the bulk field near 353\,K, with a
    broad warmer central band and cooler outer edges at the same average
    temperature as Case~1}
    \label{fig:armcore_qualcomm_tmap}
  \end{subfigure}

  \begin{subfigure}[b]{0.34\textwidth}
    \centering
    \includegraphics[width=\linewidth]{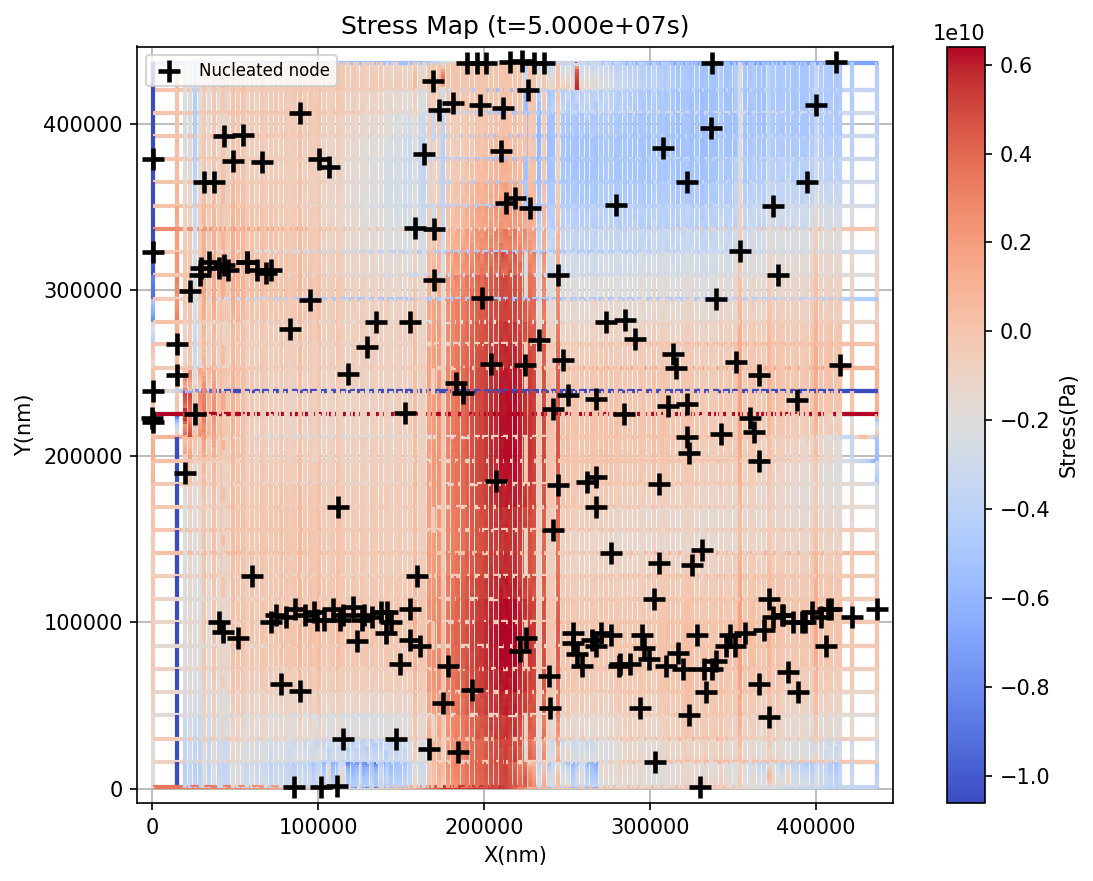}
    \caption{\small 2D stress map and nucleation sites (Case 1): 206 nucleated
    trees at $5\times10^7$\,s with a strong central tensile band}
    \label{fig:armcore_baseline_stress}
  \end{subfigure}%
  \hspace{0.04\textwidth}
  \begin{subfigure}[b]{0.34\textwidth}
    \centering
    \includegraphics[width=\linewidth]{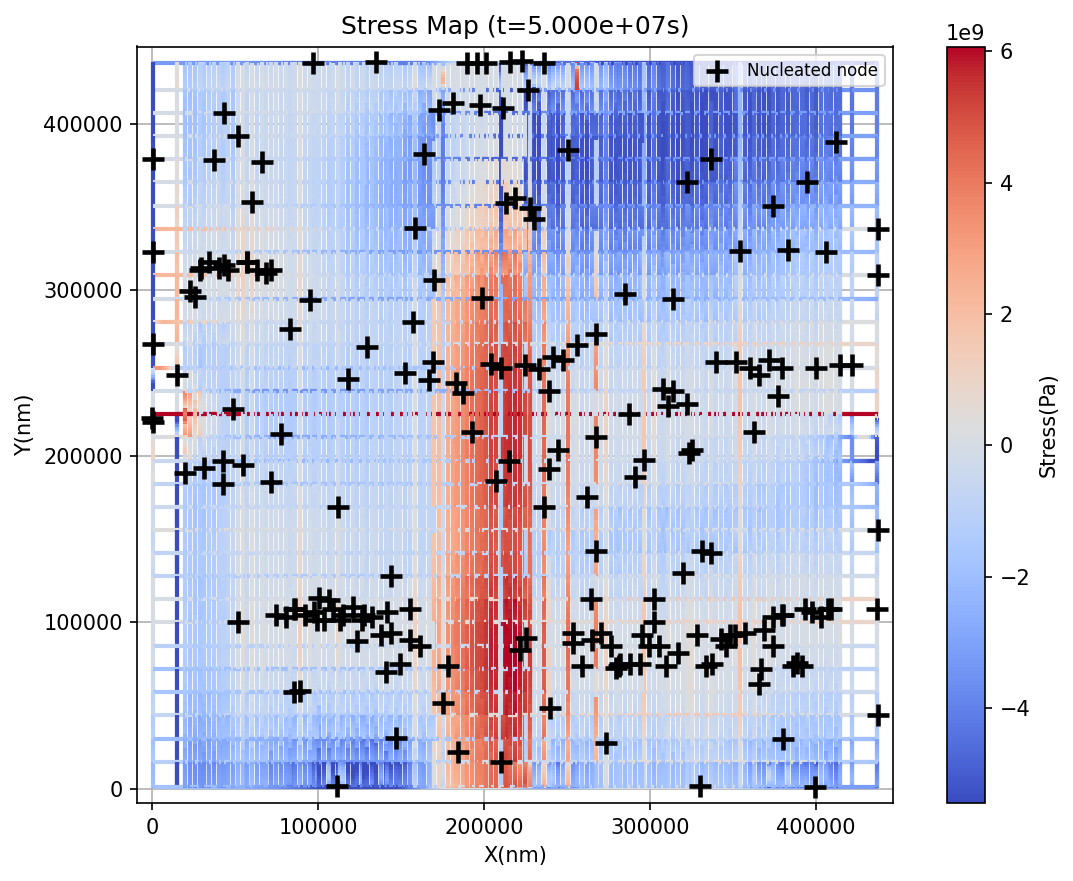}
    \caption{\small Stress map and nucleation sites (Case 2): 206 nucleated
    trees at $5\times10^7$\,s with a strong central tensile band}
    \label{fig:armcore_qualcomm_stress}
  \end{subfigure}

  \begin{subfigure}[b]{0.38\textwidth}
    \centering
    \includegraphics[width=\linewidth]{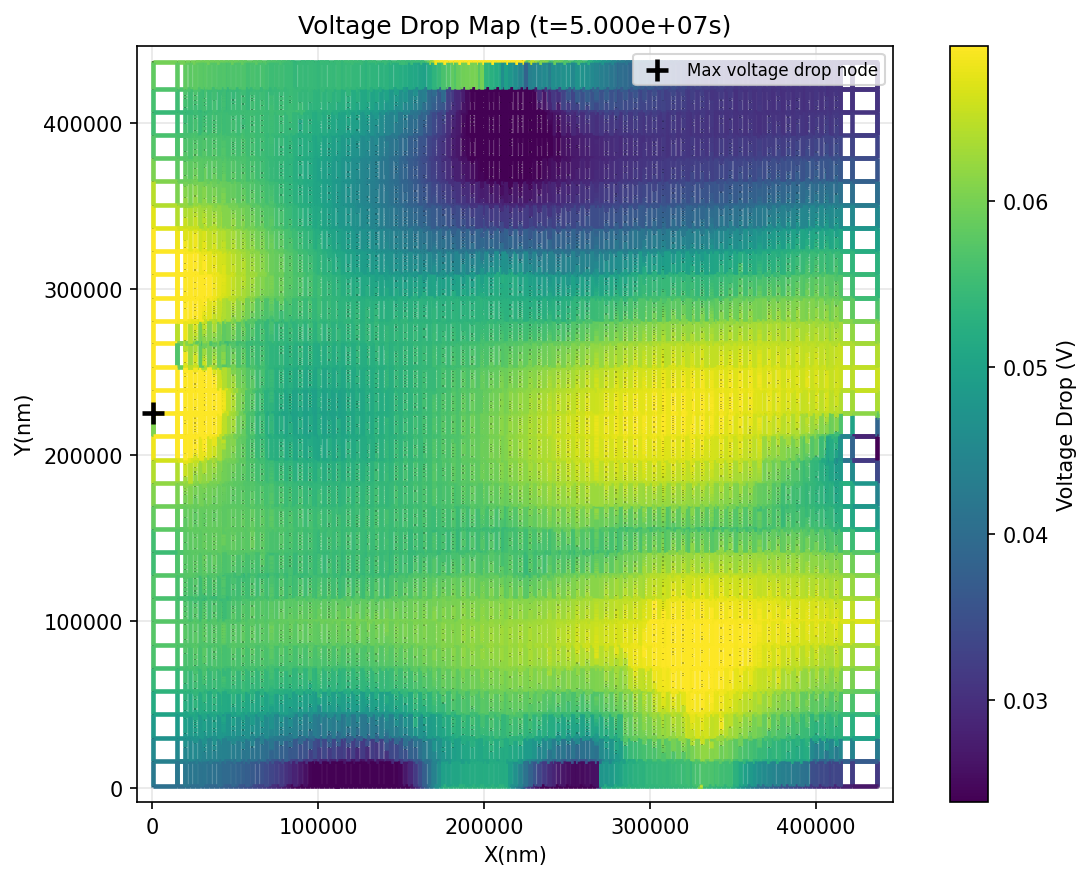}
    \caption{\small 2D IR-drop map at $t=5\times10^7$\,s (Case 1):
    max IR-drop 22.21\% with the max-drop node near the left boundary}
    \label{fig:armcore_baseline_vdrop}
  \end{subfigure}%
  \hspace{0.04\textwidth}
  \begin{subfigure}[b]{0.38\textwidth}
    \centering
    \includegraphics[width=\linewidth]{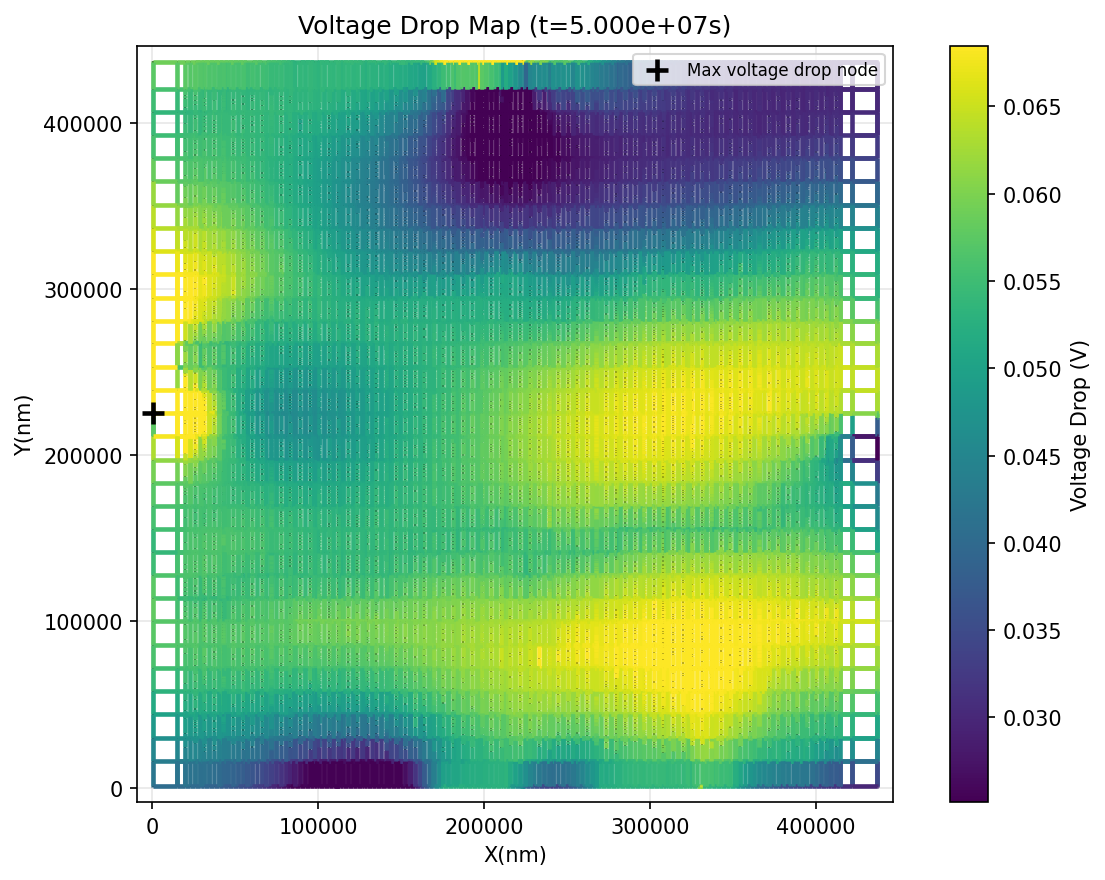}
    \caption{\small 2D IR-drop map at $t=5\times10^7$\,s (Case 2):
    max IR-drop 21.10\% with the max-drop node near the left boundary}
    \label{fig:armcore_qualcomm_vdrop}
  \end{subfigure}

  \caption{\small Temperature maps, stress distributions, and IR-drop maps
  for the ARM Cortex-A logic core under two thermal scenarios}
  \label{fig:armcore_all}
\end{figure*}

\textbf{Case 1 — Baseline external temperature map.}
The first scenario uses an updated baseline spatial temperature distribution
provided externally as input to \textit{EMSpice~3}. The resulting per-node
temperature distribution is shown in Fig.~\ref{fig:armcore_baseline_tmap}. The
rendered color scale emphasizes the bulk field near the ambient baseline, with
warmer regions concentrated in the lower-left and lower-right portions of the
die and the coolest region appearing in the upper-right corner. The full
coupled run reaches a maximum temperature of about 385.4\,K because localized
wire-segment self-heating is superimposed on the external map.

As a diagnostic, steady-state screening reports 205 transient-follow-up trees
and 3 immortal trees. In the full transient simulation used to generate
Fig.~\ref{fig:armcore_all}, 206 of the 208 trees nucleate within the simulation
horizon, confirming that nearly the entire network is vulnerable under this
thermal field. The stress map in Fig.~\ref{fig:armcore_baseline_stress} shows a
strong tensile band through the central region together with widespread
nucleation across the design. The initial maximum IR-drop is 8.85\% of
$V_{src}$; it rises to 22.21\% by the end of the simulation. The 10\%
IR-drop failure threshold is crossed at approximately $1.05\times10^7$\,s
($\approx$4.0 months), as shown in Fig.~\ref{fig:armcore_baseline_vdrop},
where the maximum-drop node lies near the left boundary of the grid.

\textbf{Case 2 — Qualcomm measured temperature map.}
The actual per-node temperature distribution derived from the Qualcomm
SM6225 measured map is shown in Fig.~\ref{fig:armcore_qualcomm_tmap}.
The rendered color scale again emphasizes the bulk field near the 353\,K
baseline, but the warmer region now forms a broad central band spanning much
of the middle of the die, while the outer edges remain cooler. This is a
different spatial pattern from Case~1, whose warmer regions are split between
the lower-left and lower-right portions of the die. In the coupled simulation,
localized self-heating again raises the maximum wire temperature to about
385.4\,K. Steady-state screening again reports 205 transient-follow-up trees
and 3 immortal trees, and 206 of the 208 trees nucleate in the full transient
simulation (Fig.~\ref{fig:armcore_qualcomm_stress}). The initial maximum IR-drop is 8.85\% of $V_{src}$; it rises to 21.10\% by the end of the simulation.
The 10\% failure threshold is crossed at $1.00\times10^7$\,s
($\approx$
3.9 months) (Fig.~\ref{fig:armcore_qualcomm_vdrop}).

\subsubsection{Joule-Heating Effects on Wire Temperature}
\begin{figure}[!t]
  \centering
  \includegraphics[width=0.82\linewidth]{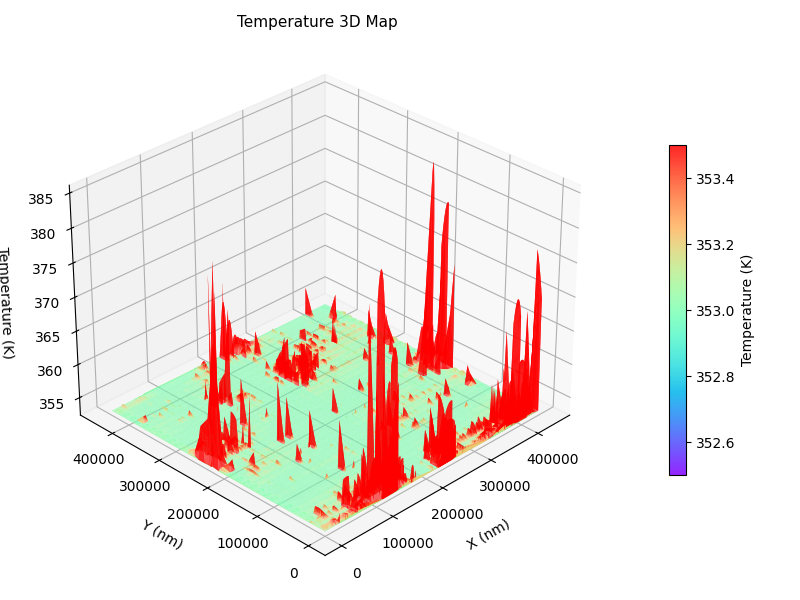}
  \caption{\small 3D wire-level temperature distribution of the ARMcore power
  grid under uniform 353.0\,K ambient temperature with Joule heating only and
  no external thermal-map input. The global field stays near the ambient
  baseline, while localized hot segments produce sharp thermal peaks over the
  interconnect network.}
  \label{fig:armcore_tmonly_3d}
\end{figure}

Fig.~\ref{fig:armcore_tmonly_3d} shows a separate ARMcore run performed with
Joule heating enabled only, without any external thermal-map input, under a
uniform 353.0\,K ambient temperature. Most of the grid remains close to the
ambient baseline, while localized sharp peaks appear along the most heavily
loaded wire segments. Rather than forming one broad die-scale hotspot, the
heating is concentrated on specific interconnect paths. The many small round
bumps distributed across the map correspond to segment-scale temperature
variation within individual wire segments, capturing localized self-heating
caused by current concentration and path resistance differences.

\begin{figure}[!t]
  \centering
  \begin{subfigure}[b]{0.47\columnwidth}
    \centering
    \includegraphics[width=\linewidth]{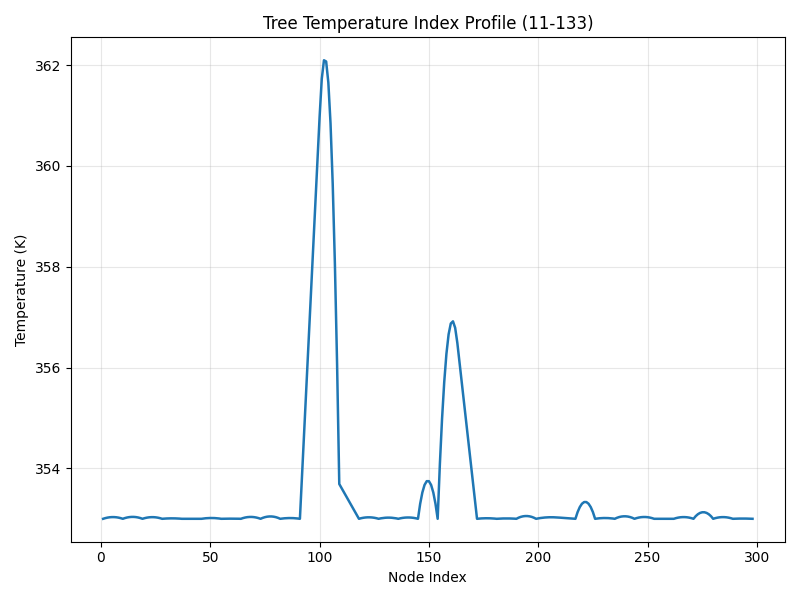}
    \caption{\small Tree 11-133}
    \label{fig:armcore_tmonly_11_133}
  \end{subfigure}%
  \hfill
  \begin{subfigure}[b]{0.47\columnwidth}
    \centering
    \includegraphics[width=\linewidth]{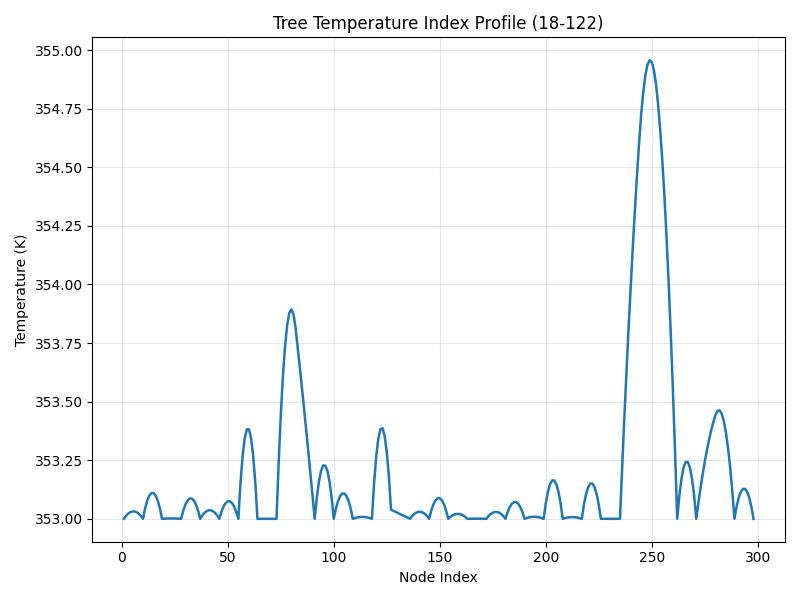}
    \caption{\small Tree 18-122}
    \label{fig:armcore_tmonly_18_122}
  \end{subfigure}

  \vspace{0.5em}

  \begin{subfigure}[b]{0.47\columnwidth}
    \centering
    \includegraphics[width=\linewidth]{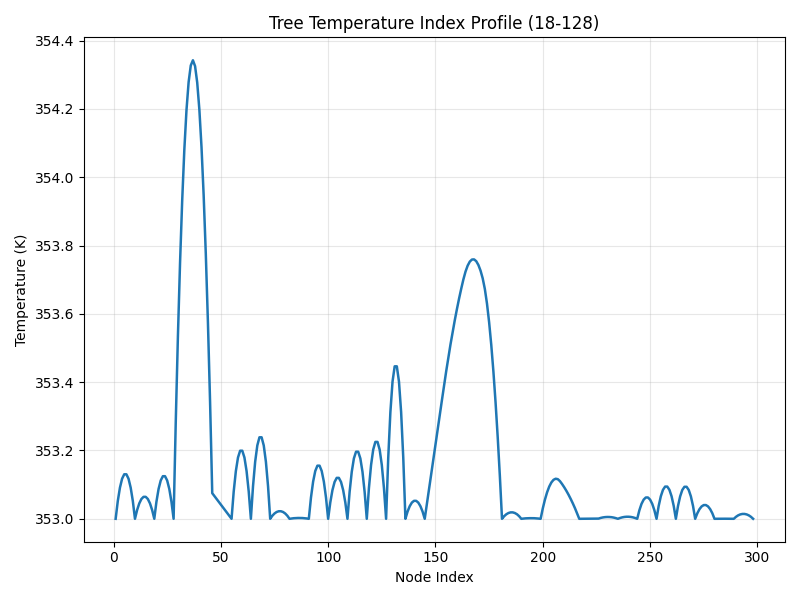}
    \caption{\small Tree 18-128}
    \label{fig:armcore_tmonly_18_128}
  \end{subfigure}%
  \hfill
  \begin{subfigure}[b]{0.47\columnwidth}
    \centering
    \includegraphics[width=\linewidth]{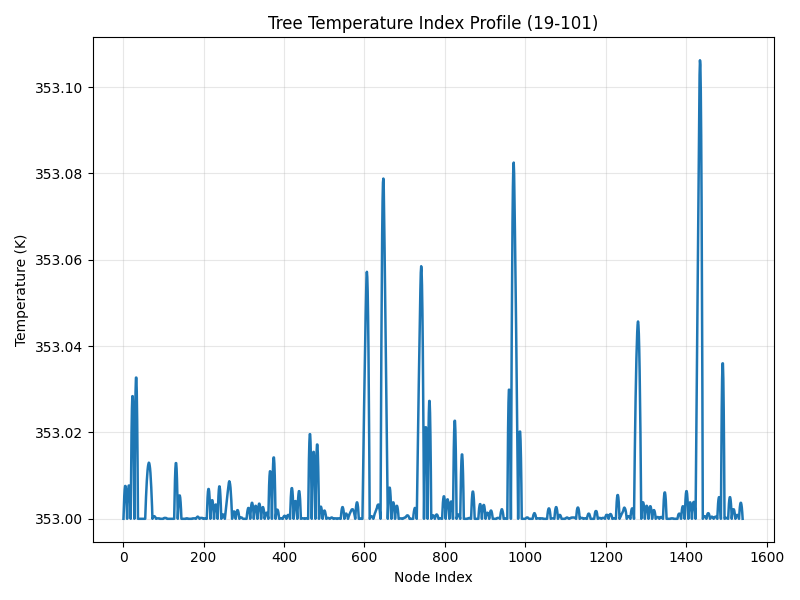}
    \caption{\small Tree 19-101}
    \label{fig:armcore_tmonly_19_101}
  \end{subfigure}
  \caption{\small Representative node-index temperature profiles for four
  ARMcore trees under uniform 353.0\,K ambient temperature with Joule heating
  only and no external thermal-map input. The profiles highlight localized
  intra-tree hot spots superimposed on a near-uniform 353\,K baseline.}
  \label{fig:armcore_tmonly_profiles}
\end{figure}

Fig.~\ref{fig:armcore_tmonly_profiles} gives representative node-index
temperature profiles for four trees from the same Joule-heating-only run under
the same 353.0\,K uniform ambient and without any external thermal map. Tree
11-133 exhibits the strongest localized hotspot among the four profiles, with a
sharp dominant spike and a secondary heated region along the same tree. Trees
18-122 and 18-128 show more moderate but still structured local peaks
superimposed on a near-353\,K baseline, indicating distinct self-heating zones
within otherwise cool trees. In contrast, tree 19-101 remains nearly flat with
only very small excursions above baseline, showing that some trees experience
minimal self-heating despite the same ambient condition. These rounded local
peaks visualize thermal spreading within individual wire segments rather than a
die-wide ambient heating effect.

Case~1 produces 206 nucleated trees and reaches the 10\% IR-drop failure
threshold at $1.05\times10^7$\,s, while the Qualcomm map also produces 206
nucleated trees and reaches failure at $1.00\times10^7$\,s. Unlike the
RISC-V core, where the thermal profile changed both the mortality regime and
the dominant failure locations more dramatically, the ARMcore remains in the
same nearly fully mortal regime under both 353\,K external temperature maps.
For this design, changing the map primarily redistributes hotspot locations
and local stress patterns, while the overall IR-drop failure time remains
clustered near $10^7$\,s.

Tab.~\ref{tab:armcore_thermal} shows that the ARMcore remains close to a
fully mortal regime across seven profiles: the mortal-tree count is 207 for
the uniform 353\,K case and 206 for all spatially varying thermal profiles.
This weak dependence on temperature profile occurs because the M1-layer current
densities push nearly every tree far above the nucleation threshold. The TTF
values also remain tightly grouped, from $1.00\times10^7$\,s to
$1.05\times10^7$\,s, even though the maximum wire temperature ranges from
341.81\,K to 405.59\,K across the nonuniform maps. At 353\,K average
temperature, the Joule-heating and Qualcomm profiles have nearly identical
maximum temperatures of 385.44\,K and produce similar fully degraded behavior.
Thus, for ARMcore, the thermal map changes the local hotspot and stress
distribution, but the aggregate failure metric is dominated by the already
deeply mortal current-density regime.

\begin{table}[!htb]
  \centering
  \caption{\small ARM Cortex-A logic core EM results under different thermal
  profiles (208 trees total). 
  TTF of the power grid is the time at which the network-wide IR-drop exceeds the 10\% failure threshold.}
  \label{tab:armcore_thermal}
  \footnotesize
  \setlength{\tabcolsep}{3pt}
  \begin{tabular}{c|cccc}
    \hline\hline
    Thermal Profile &
      \shortstack{Avg.\\Temp (K)} &
      \shortstack{Max\\Temp (K)} &
      \shortstack{Mortal\\Trees} &
      TTF (s) \\
    \hline\hline
    Uniform 353\,K                    & 353.00 & 353.00 & 207 & $1.05\times10^7$ \\
    Joule-heating (avg.\ 320\,K) & 320.42 & 352.71 & 206 & $1.04\times10^7$ \\
    Joule-heating (avg.\ 353\,K) & 353.14 & 385.44 & 206 & $1.05\times10^7$ \\
    Joule-heating (avg.\ 373\,K) & 373.28 & 405.59 & 206 & $1.05\times10^7$ \\
    Qualcomm map (avg.\ 309\,K)  & 309.55 & 341.81 & 206 & $1.01\times10^7$ \\
    Qualcomm map (avg.\ 353\,K)  & 353.22 & 385.44 & 206 & $1.00\times10^7$ \\
    Qualcomm map (avg.\ 373\,K)  & 373.40 & 405.59 & 206 & $1.00\times10^7$ \\
    \hline\hline
  \end{tabular}
  \vspace{-0.1in}
\end{table}

\subsection{Cross-Design Runtime and Accuracy Evaluation}

\begin{table*}[!htb]
  \centering
  \caption{\small Performance comparison of \textit{EMSpice~3} across benchmark designs. A TTF entry of $>5.0\times10^7$\,s indicates that the 10\% IR-drop failure threshold was not reached within the simulated time window. The Krylov metric error is measured relative to the default non-Krylov FDTD analysis, using TTF when failure occurs and the final IR-drop otherwise.}
  \label{tab:comparison}

\renewcommand{\arraystretch}{1.3} %
\footnotesize
\setlength{\tabcolsep}{3pt}
\begin{tabular}{@{}c|ccccccccccc@{}}
\hline\hline
\makecell[c]{Design\\Name} &

\makecell[c]{No.\\Trees} &
\makecell[c]{Max Branch.\\per Tree} &
\makecell[c]{Max Nodes\\per Tree} &
\makecell[c]{Init.\ IR\\Drop (\%)} &
\makecell[c]{Final IR\\Drop (\%)} &
\makecell[c]{TTF (s)} &
\makecell[c]{Mortal\\Seg. Count} &
\makecell[c]{Baseline\\CPU (s)} &
\makecell[c]{Krylov\\CPU (s)} &
\makecell[c]{Speedup} &
\makecell[c]{Krylov Metric\\Error (\%)} \\
\hline\hline
AES\_new          & 97  & 69  & 3450  & 6.14  & 6.86  & $>5.0\times10^7$ &   688 &  4.56 &  3.76 & 1.21 & 
$<0.05$ \\
armcore\_pad      & 68  & 34  & 1700  & 0.29  & 0.29  & $>5.0\times10^7$ &     0 &  2.35 &  1.99 & 1.18 & $<0.05$ \\
JPEG\_new         & 178 & 126 & 6300  & 6.56  & 6.77  & $>5.0\times10^7$ &  1821 & 16.31 & 12.90 & 1.26 & $<0.05$ \\
armcore\_logic    & 208 & 218 & 10900 & 8.85   & 22.21 & $1.05\times10^7$ &   206 & 40.60 & 31.48 & 1.29 & $<0.05$ \\
dual\_ram         & 55  & 29  & 1450  & 0.09  & 0.10  & $>5.0\times10^7$ &   179 &  2.32 &  1.55 & 1.50 & $<0.05$ \\
risc\_core        & 186 & 115 & 5750  & 6.19  & 29.64 & $2.47\times10^7$ &    18 &  5.40 &  4.45 & 1.21 & $<0.05$ \\
\hline\hline
\end{tabular}

\end{table*}

To further evaluate computational cost and reduction effectiveness across
different network structures, we tested the framework on a broader set of
benchmark designs. Tab.~\ref{tab:comparison} summarizes the key statistics and runtime results
for six designs, including the two case studies above. 
\textit{AES\_new} is an AES encoder design, \textit{JPEG\_new} is a JPEG encoder design, \textit{armcore\_logic} is an ARM Cortex-A logic core design, \textit{armcore\_pad} is an ARM Cortex-A pad design, and \textit{dual\_ram} is a dual-port RAM design, and \textit{risc\_core} is a RISC-V core design. 
The PG networks of the six designs were obtained from the Synopsys Fusion Compiler. All designs were synthesized and placed-and-routed using the SAED32
(Synopsys 32/28nm Generic) library~\cite{SynopsysGenericLibrary}. 

For each design, we
report the number of trees, the maximum number of branches and nodes per tree,
the initial and final maximum IR-drop (as a percentage of $V_{src}$), the corresponding IR-drop
time to failure (TTF), the mortal count, and the wall-clock CPU time (in
seconds) with and without the Krylov-subspace acceleration described in
Section~\ref{subsec:rakrylov}. For the two case-study designs, the listed TTF
and mortal-count values correspond to the 353\,K Joule-heating cases reported
in Tabs.~\ref{tab:risc_thermal} and~\ref{tab:armcore_thermal}, so that the
cross-design comparison remains tied to a single operating point. The Krylov
metric error uses the default non-Krylov FDTD simulation as the reference: it
compares TTF for designs that cross the 10\% IR-drop threshold and final
maximum IR-drop for designs that do not fail within the simulation horizon.

Several observations follow from Table~\ref{tab:comparison}.
The Krylov-based solver consistently reduces runtime, with speedups from
$1.18\times$ (\textit{armcore\_pad}) to $1.50\times$ (\textit{dual\_ram}),
while matching the default non-Krylov FDTD reference with 
less than 0.05\% reported metric error for all six benchmarks.
The final IR-drop exceeds the initial value for all designs with mortal
interconnects. Notably, \textit{risc\_core} shows a dramatic increase from
6.19\% to 29.64\% despite only 18 mortal trees in the corresponding
Table~\ref{tab:risc_thermal} operating point, underscoring that a few
high-current trees can dominate IR-drop degradation.
\textit{armcore\_pad} records zero mortal interconnects, validating that
temperature-aware screening efficiently prunes immortal trees before
the transient FDTD stage.
Runtimes scale with design complexity: the largest design,
\textit{armcore\_logic} (208 trees, up to 10,900 nodes), completes in
about 31.5\,s with Krylov acceleration, while the smallest,
\textit{dual\_ram} (55 trees), finishes in under 1.6\,s, confirming
that the coupled analysis remains computationally tractable at the
full-chip scale.

\subsection{Monte Carlo EM Reliability Analysis}

To quantify the statistical variability of EM-induced time to failure
(TTF) arising from manufacturing process variations, we perform Monte
Carlo (MC) simulations on both the RISC-V core and the ARM Cortex-A
logic core designs.  In each MC run, the 
EM diffusivity~$\kappa(x)$ and critical stress parameters are randomly perturbed with a
coefficient of variation (CoV) of 20\%.  For each
design, 100 independent MC samples are drawn, and
the full \textit{EMSpice~3} pipeline—immortality screening, transient FDTD
simulation, and coupled IR-drop analysis—is re-executed for each sample.
The RISC-V core uses the Joule-heating temperature map at 353\,K; the ARM
Cortex-A logic core uses the corresponding 353\,K Joule-heating thermal
condition derived from the first input thermal-map case
(Fig.~\ref{fig:armcore_baseline_tmap}).  The 10\% IR-drop threshold is used to define
the TTF for each run.  A run in which the IR-drop never exceeds the threshold
within the $5\times10^7$\,s simulation horizon is recorded as
\emph{censored} (right-censored).

\begin{figure}[!htb]
  \centering
  \includegraphics[width=0.60\linewidth]{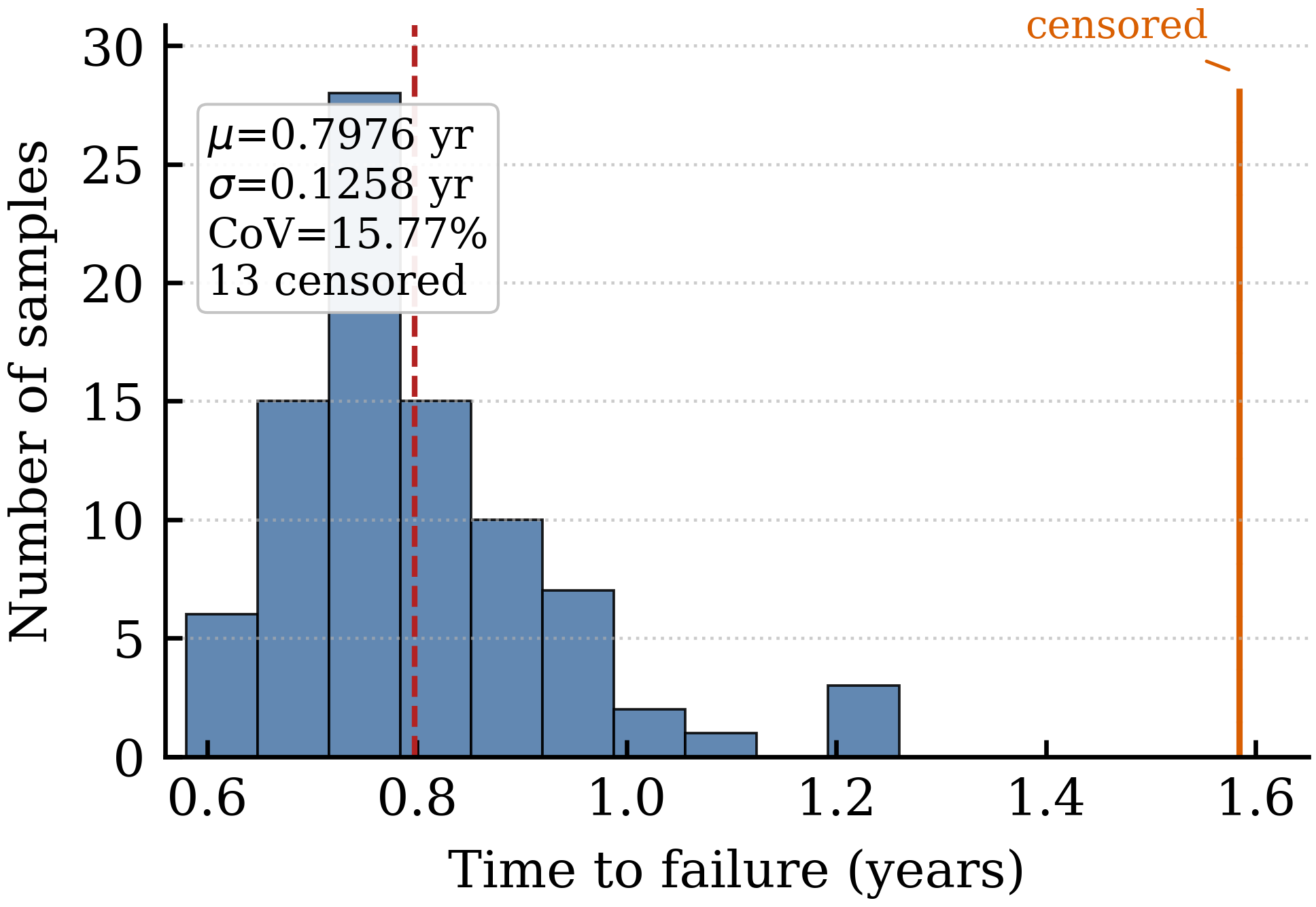}
  \caption{\small Monte Carlo TTF distribution for the RISC-V core
  (100 runs; 20\% variation in $\kappa(x)$ and critical stress).
  Uncensored failures: 87 runs, mean TTF\,=\,0.7976\,years,
  $\sigma$\,=\,0.1258\,years; 13 runs are right-censored at
  $5\times10^7$\,s.}
  \label{fig:mc_risc}
\end{figure}

\textbf{RISC-V Core.}
Fig.~\ref{fig:mc_risc} shows the TTF histogram for the RISC-V core
across 100 MC runs.  Of the 100 successful runs, 87 result in IR-drop
failure within the simulation window and 13 are right-censored at
$5\times10^7$\,s.  The mean TTF among the 87 observed failures is
$0.7976$\,years ($\approx 2.517\times10^{7}$\,s) with a standard deviation of
$0.1258$\,years ($\approx 3.969\times10^{6}$\,s), yielding CoV of 15.77\%.
These statistics are computed only from the
uncensored failure times; the censored samples are shown separately at the
simulation horizon in Fig.~\ref{fig:mc_risc}.  The uncensored failures span
0.6046--1.2298\,years, while the censored samples indicate that some
parameter draws remain below the IR-drop failure threshold throughout the
1.5844-year simulation window.  This spread underscores that deterministic
single-point EM analysis can substantially over- or under-estimate the actual
failure time for designs where a small number of high-current trees dominate
the failure mechanism.

\begin{figure}[!htb]
  \centering
  \includegraphics[width=0.60\linewidth]{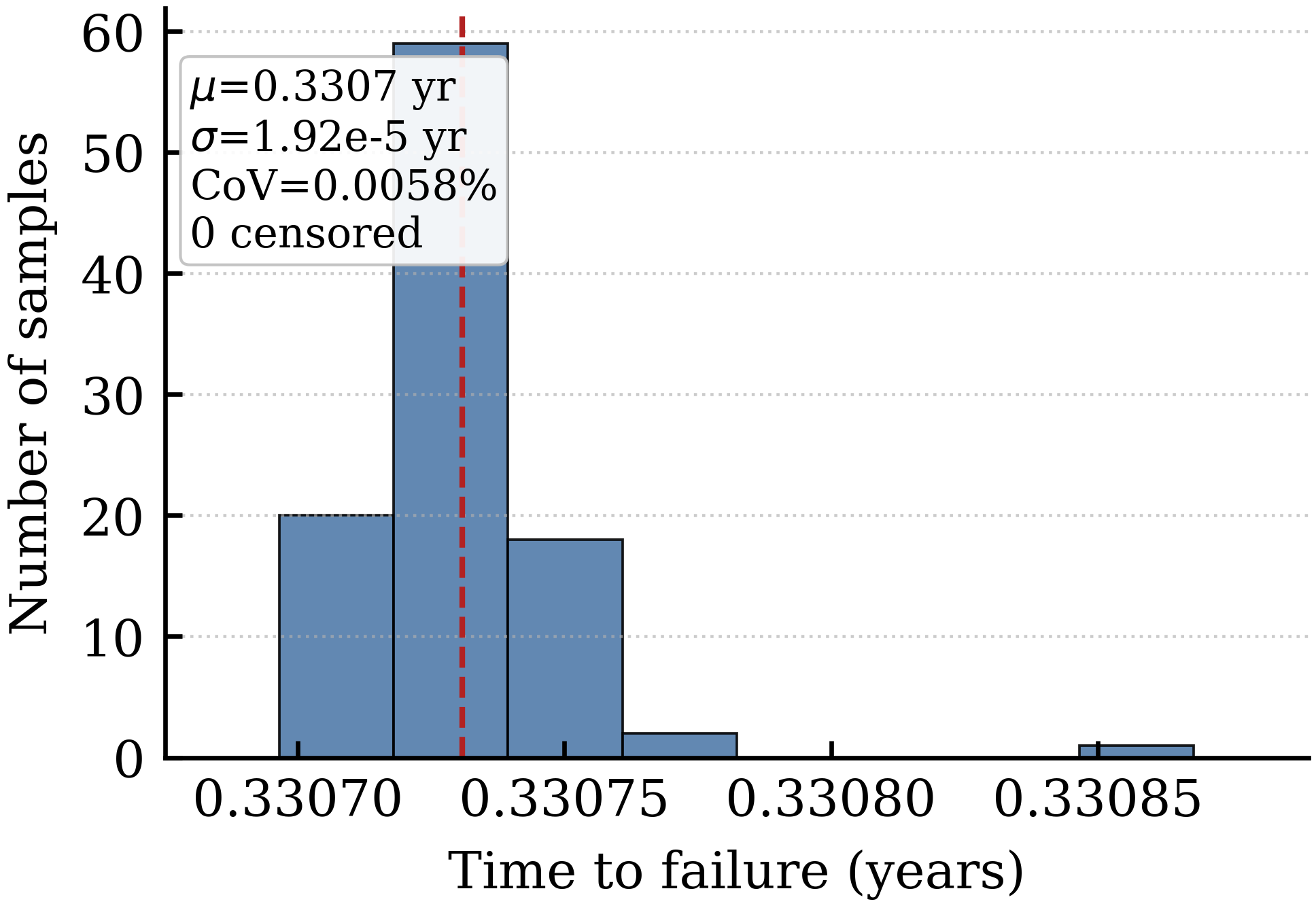}
  \caption{\small Monte Carlo TTF distribution for the ARM Cortex-A
  logic core (100 runs; 20\% variation in $\kappa(x)$ and critical
  stress).  Mean TTF\,=\,0.3307\,years,
  $\sigma$\,=\,$1.92\times10^{-5}$\,years; 0 censored runs.}
  \label{fig:mc_armcore}
\end{figure}

\textbf{ARM Cortex-A Logic Core.}
Fig.~\ref{fig:mc_armcore} shows the corresponding TTF histogram for the
ARM Cortex-A logic core under the same 353\,K Joule-heating thermal condition.
All 100 runs result in IR-drop failure (zero censored).  The mean TTF is
$0.3307$\,years ($\approx 1.0437\times10^{7}$\,s) with a standard deviation
of $1.92\times10^{-5}$\,years ($\approx 606.1$\,s), corresponding to a
CoV of only 0.0058\%.
Virtually all samples collapse
onto a single narrow bin between approximately 0.3307 and 0.3309\,years.  This
near-deterministic behavior is
consistent with the ARM Cortex-A case study results in Tab.~\ref{tab:armcore_thermal}: the
ARMcore design has 206 mortal trees under the 353\,K Joule-heating thermal
profile. Its M1-layer current
densities far exceed the nucleation threshold.  When the overwhelming
majority of trees are deeply mortal, random perturbations in
EM diffusivity~$\kappa(x)$ and critical stress
have negligible effect on which trees fail or when, because the design
is already operating far into the failure regime.  The TTF is therefore
dominated by the deterministic stress buildup rate rather than by
stochastic parameter variations.

\textbf{Comparison and Discussion.}
The contrasting results reveal design-dependent variation sensitivity.
The RISC-V core, with only 18 mortal trees, exhibits a 15.77\% TTF
CoV because failure hinges on marginally mortal trees
that small perturbations in $\kappa(x)$ and critical stress can push across
the nucleation threshold. The ARMcore, where 206 trees are deeply
mortal, shows virtually no TTF sensitivity to the same perturbations; there,
the TTF is dominated by deterministic stress buildup. This dichotomy suggests
that designs with few marginally mortal trees benefit most from statistical
analysis, whereas deeply mortal designs can be much less sensitive to the same
variation sources. Together, these results confirm that both the spatial
temperature structure and the margin of individual trees relative to the
nucleation threshold must be characterized for predictive full-chip EM
reliability analysis.

\section{Conclusions}
\label{sec:conclusion}

  We presented \textit{EMSpice~3}, a full-chip multiphysics framework for coupled EM, TM, and IR-drop analysis of power-grid networks.
  The results show that spatial thermal structure has a first-order impact on predicted reliability: equal-average thermal profiles can change the number of mortal trees, shift the void nucleation location and degraded branch inside a critical tree, and determine whether the 10\% IR-drop failure criterion is reached within the simulation horizon. 
  The RISC-V results further show that the relationship between average temperature and TTF can be non-monotonic, because a higher-average-temperature profile can be less damaging when its hotspots are not aligned with the critical current paths. 
  Monte Carlo analysis also shows that process-variation sensitivity is strongly design-dependent: the RISC-V core exhibits about 15.8\% TTF coefficient of variation under 20\% variation in 
  EM diffusivity~$\kappa(x)$ and critical stress, while the ARM Cortex-A logic core shows only 0.0058\%.
  Together with the consistent $1.18\times$--$1.50\times$ Krylov speedup and 
  sub-0.05\% reported metric error relative to the default non-Krylov FDTD analysis, these results indicate that predictive chip-scale EM analysis requires coupled treatment of spatial thermal fields, resistance feedback, and stochastic material variation.


\section*{Acknowledgments}
The authors would like to thank the anonymous reviewers of the ISVLSI 2024 conference version for their valuable feedback.
Generative AI tools, specifically ChatGPT and Claude Code, were used to assist with editing and proofreading of this manuscript.

\bibliographystyle{IEEEtran}

\bibliography{../../bib/security,../../bib/emergingtech,../../bib/thermal_power,../../bib/mscad_pub,../../bib/interconnect,../../bib/stochastic,../../bib/simulation,../../bib/modeling,../../bib/reduction,../../bib/misc,../../bib/architecture,../../bib/reliability,../../bib/reliability_papers,../../bib/machine_learning,../../bib/neural_network}

\end{document}